# Numerical simulation of surface waves instability on a discrete grid.

Alexander O. Korotkevich[a,b,∗], Alexander I. Dyachenko[b,e], Vladimir E. Zakharov[c,d,e,f,b]

[a]*Department of Mathematics & Statistics, The University of New Mexico, MSC01 1115, 1 University of New Mexico, Albuquerque, New Mexico, 87131-0001, USA*
[b]*L. D. Landau Institute for Theoretical Physics, 2 Kosygin Str., Moscow, 119334, Russian Federation*
[c]*University of Arizona, Department of Mathematics, Tucson, AZ 85721, USA*
[d]*P. N. Lebedev Physical Institute, 53 Leninskiy prospekt, Moscow, 119334, Russian Federation*
[e]*Laboratory of Nonlinear Wave Processes, Novosibirsk State University, Novosibirsk, Russian Federation*
[f]*Waves and Solitons LLC,738 W. Sereno Dr., Gilbert, AZ, USA*

**Abstract**

We performed full-scale numerical simulation of instability of weakly nonlinear waves on the surface of deep fluid. We show that the instability development leads to chaotization and formation of wave turbulence.

Instability of both propagating and standing waves were studied. We studied separately pure capillary wave unstable due to three-wave interactions and pure gravity waves unstable due to four-wave interactions. The theoretical description of instabilities in all cases is included into the article. The numerical algorithm used in these and many other previous simulations performed by authors is described in details.

*Keywords:*
*PACS:*

∗Corresponding author
   *Email address:* `alexkor@math.unm.edu` (Alexander O. Korotkevich)

# 1. Introduction.

Stationary propagating waves on the surface of deep heavy ideal fluid are known since the middle of nineteenth century. Stokes (see, for instance [1]) in 1847 found the solution of the Euler equation in the form of trigonometric series. For the shape of surface $\eta(x,t)$ he obtained

$$\eta(x,t) = a\left[\cos(kx - \omega t) + \frac{1}{2}\mu\cos\{2(kx - \omega t)\} + \frac{3}{8}\mu^2\cos\{3(kx - \omega t)\} + \ldots\right]. \quad (1)$$

Here we introduced steepness $\mu$ and frequency $\omega$

$$\mu = ka, \quad \omega = \sqrt{gk}\left(1 + \frac{1}{2}\mu^2 + \frac{1}{8}\mu^4 + \ldots\right). \quad (2)$$

Stokes found two algorithms for calculation of all terms in series (1) and (2) (See Sretenskii [2]). Convergence of these series has been proven by Nekrasov [3, 4] in 1921. Another proof was found by Levi-Chivita [5].

It is known since 1965 [6] that stationary waves on surface of deep water are unstable. The theory of instability [7, 8, 9, 10] was developed for waves of small amplitude in the limit $\mu \to 0$. A history of this question is described in the article [11]. In the present paper we study instability of stationary waves numerically by direct solution of the Euler equation describing a potential flow of ideal fluid with free surface. This approach has two following important advantages. Firstly, by numerical simulation we can study waves with finite amplitude. This is mostly a program for future, in this paper we shall study only waves of small amplitude. The second advantage is not less serious. By the use of numerical simulation we can study not only linear, but also nonlinear stage of instability development. Even in integrable systems like the NLSE, analytical study of the monochromatic wave is a very nonlinear problem which could be solved only by methods of algebraic geometry. In more realistic models development of a nonlinear theory of modulation waves instability is a hopeless problem. In a long run we have to expect that the instability will lead to formation of a stochastic wave field described by a kinetic equation for squared wave amplitudes and formation of Kolmogorov-Zakharov (KZ) spectra, governed by the energy flux in high wave numbers [12].

The article is organized as follows. Chapters 2 and 3 are devoted to analytical theory of stability of weakly nonlinear stationary waves. To develop this theory we use Hamiltonian formalism as the most compact and suitable. We start with presenting the Euler equation of ideal fluid with free surface in the Hamiltonian form. Surface tension is included in the Hamiltonian. In the presence of surface tension the dispersion relation is

$$\omega_k = \sqrt{gk + \sigma k^3},$$

here $\sigma$ – surface tension coefficient.

Wave vectors of small-amplitude stationary waves are solutions of equation

$$\omega_k = ck. \qquad (3)$$

This equation has two solutions

$$k_{1,2} = \frac{c \pm \sqrt{c^4 - 4g\sigma}}{2\sigma}, \qquad (4)$$

if $c > c_0$, where $c_0 = (4g\sigma)^{1/4}$. For water $c_0 \simeq 12 cm/sec$. In a generic case $c\ c_0$ stationary waves comprise a complicated four-parameter family. However, in the limiting case $c \gg c_0$ one can split it in two periodic families of "pure gravitational" and "pure capillary" waves.

The Stokes wave is "pure gravitational". Now $k_1 = g/c^2$ and capillary effects could be neglected. In the "pure capillary" case $k_2 = c^2/\sigma$ and effects of gravity can be neglected. All stationary waves on the surface of deep fluid are unstable. However, instability of short capillary waves and long gravity waves have different flavor. They described by different "efficient Hamiltonians". The case of "pure capillary" waves is the simplest. The instability can be studied if framework of the Hamiltonian contains only quadratic and cubic terms. This is a subject of Chapter 2. A situation is more complicated for gravitational waves. In this case terms of fourth order must be included in the Hamiltonian. Then one have to exclude the cubic terms by a proper conformal transformation. As a result we get so-called "Zakharov equation" [10]. In the framework of this equation the problem of the Stokes wave stability can be solved exactly. This is a subject of Chapter 3.

In Chapter 4 we give a detailed description of the numerical code which we used for solution of the Hamiltonian Euler equation. This code was used in many papers but never was described in details [13, 14, 15, 16, 17, 18, 19, 20]. We should stress that in our numerical experiments we worked with the Euler equation written in "natural variables". These equation are not as good for direct analytical study as they good for implementation of numerical method. The structure of nonlinear parts of the Hamiltonian in "natural variables" is relatively simple and using the standard Fast Fourier Transform (FFT) method is quite possible.

In Chapter 5 we present our results on modeling of capillary wave instability. We show that an initial stage of instability is described pretty well by the linear analytical theory. Further development of instability demonstrates appearing of "secondary instabilities" and a tendency to formation of a chaotic wave field which should be described by statistical methods.

In Chapter 6 we study instability of the Stokes wave. We show that this instability is mostly "modulational". In other words the wave remains quasi-monochromatic for a long time after development of the instability.

Finally, in Chapter 7 we present first results on development of the standing waves instability. We show that this instability leads to fast isotropization of the wave field. This mechanism can be used in experiments for generation of isotropic wave field.

## 2. Theory of decay instability

In this section we develop the simplest version of the theory of stationary waves instability. This simple theory is applicable if triple-wave nonlinear processes governed by the resonant conditions

$$\omega_k = \omega_{k_1} + \omega_{k_2}, \tag{5}$$
$$\vec{k} = \vec{k}_1 + \vec{k}_2. \tag{6}$$

are permitted. But before we briefly describe how the theory of surface waves can be embedded into the general Hamiltonian theory of nonlinear waves.

Suppose that ideal incompressible fluid fills the space $-\infty < z < \eta(\vec{r}, t)$, here $\vec{r} = (x, y)$ —two dimensional vector. A flow is potential $\vec{v} = \nabla \Phi$, hence hydrodynamical potential $\Phi$ satisfies the Laplace equation

$$\Delta \Phi = 0. \tag{7}$$

Let us define $\psi = \Phi|_{z=\eta}$ and impose a natural boundary condition $Phi_z \to 0$ at $z \to -\infty$. It is known [7] that $\eta(\vec{r}, t)$ and $\psi(\vec{r}, t)$ are canonically conjugated variables satisfying evolutionary equations

$$\frac{\partial \eta}{\partial t} = \frac{\delta H}{\delta \psi}, \quad \frac{\partial \psi}{\partial t} = -\frac{\delta H}{\delta \eta}. \tag{8}$$

Here $H = T + U$ — total energy of the fluid, consisting of kinetic energy

$$T = \frac{1}{2} \int d^2\vec{r} \int_{-\infty}^{\eta} (\nabla \Phi)^2 dz, \tag{9}$$

and potential energy

$$U = \frac{g}{2} \int \eta^2 d^2\vec{r} + \sigma \int (\sqrt{1 + (\nabla \eta)^2} - 1) d^2\vec{r}. \tag{10}$$

The Hamiltonian $H$ in terms of $\eta$ and $\psi$ is given by the infinite series

$$H = H_0 + H_1 + H_2 + \ldots \tag{11}$$

Here

$$H_0 = \frac{1}{2} \int \left\{ \psi \hat{k} \psi + g\eta^2 + \sigma(\nabla \eta)^2 \right\} d^2\vec{r}, \tag{12}$$

$$\text{here } \hat{k}\psi = \sqrt{-\Delta}\psi,$$

$$H_1 = \frac{1}{2} \int \eta \{|\nabla \psi|^2 - (\hat{k}\psi)^2\} d^2\vec{r}, \tag{13}$$

$$H_2 = \frac{1}{2} \int \eta(\hat{k}\psi)[\hat{k}(\eta \hat{k}\psi) + \eta \Delta \psi] d^2\vec{r} + \frac{1}{2}\sigma \int (\nabla \eta^2)^2 d^2\vec{r}. \tag{14}$$

Thereafter we will neglect last term in (14).

One can perform the symmetric Fourier transform

$$\psi_{\vec{k}} = \frac{1}{2\pi} \int \psi(\vec{r}) e^{-i\vec{k}\vec{r}} d^2\vec{r}, \quad \eta_{\vec{k}} = \frac{1}{2\pi} \int \eta(\vec{r}) e^{-i\vec{k}\vec{r}} d^2\vec{r}. \tag{15}$$

This is the canonical transformation. Equations (8) now take the form

$$\frac{\partial \eta}{\partial t} = \frac{\delta H}{\delta \psi^*}, \quad \frac{\partial \psi}{\partial t} = -\frac{\delta H}{\delta \eta^*}. \tag{16}$$

Now

$$\begin{aligned}
H_0 &= \frac{1}{2} \int (|k||\psi_{\vec{k}}|^2 + \sigma |k|^2 |\eta_{\vec{k}}|^2 + g|\eta_{\vec{k}}|^2) d\vec{k}, \\
H_1 &= -\frac{1}{4\pi} \int L_{\vec{k}_1 \vec{k}_2} \psi_{\vec{k}_1} \psi_{\vec{k}_2} \eta_{\vec{k}_3} \delta(\vec{k}_1 + \vec{k}_2 + \vec{k}_3) d\vec{k}_1 d\vec{k}_2 d\vec{k}_3, \\
H_2 &= \frac{1}{16\pi^2} \int M_{\vec{k}_1 \vec{k}_2 \vec{k}_3 \vec{k}_4} \psi_{\vec{k}_1} \psi_{\vec{k}_2} \eta_{\vec{k}_3} \eta_{\vec{k}_4} \delta(\vec{k}_1 + \vec{k}_2 + \vec{k}_3 + \vec{k}_4) d\vec{k}_1 d\vec{k}_2 d\vec{k}_3 d\vec{k}_4,
\end{aligned} \tag{17}$$

Here

$$\begin{aligned}
L_{\vec{k}_1 \vec{k}_2} &= (\vec{k}_1 \vec{k}_2) + |\vec{k}_1||\vec{k}_2|, \\
M_{\vec{k}_1 \vec{k}_2 \vec{k}_3 \vec{k}_4} &= |\vec{k}_1||\vec{k}_2| \left[ \frac{1}{2}(|\vec{k}_1 + \vec{k}_3| + |\vec{k}_1 + \vec{k}_4| + \right. \\
&\quad + |\vec{k}_3 + \vec{k}_2| + |\vec{k}_2 + \vec{k}_4|) - |\vec{k}_1| - |\vec{k}_2| \Big].
\end{aligned} \tag{18}$$

Equations (8) written for the Hamiltonian (13-14) read

$$\begin{aligned}
\dot{\eta} &= \hat{k}\psi - (\nabla(\eta \nabla \psi)) - \hat{k}[\eta \hat{k}\psi] + \hat{k}(\eta \hat{k}[\eta \hat{k}\psi]) + \\
&\quad + \tfrac{1}{2}\nabla^2[\eta^2 \hat{k}\psi] + \tfrac{1}{2}\hat{k}[\eta^2 \nabla^2 \psi], \\
\dot{\psi} &= \sigma \nabla^2 \eta - g\eta - \tfrac{1}{2}\left[(\nabla \psi)^2 - (\hat{k}\psi)^2\right] - [\hat{k}\psi]\hat{k}[\eta \hat{k}\psi] - [\eta \hat{k}\psi]\nabla^2 \psi.
\end{aligned} \tag{19}$$

These equations were considered for the first time in [21]. Equations (19) are basic in our numerical simulations. To develop analytical theory of stationary waves instability we use equation (16). It is interesting to note that even for cappilary waves in some cases it is worthwhile to keep cubic terms in (19) as it is shown in Appendix C.

Let us introduce complex normal variables

$$a_{\vec{k}} = \sqrt{\frac{\omega_k}{2k}} \eta_{\vec{k}} + i \sqrt{\frac{k}{2\omega_k}} \psi_{\vec{k}}. \tag{20}$$

As far as $\eta_{-\vec{k}} = \eta_{\vec{k}}^*$, $\psi_{-\vec{k}} = \psi_{\vec{k}}^*$ (because these are Fourier transforms of real functions) we have

$$\eta_{\vec{k}} = \sqrt{\frac{2k}{\omega_k}}(a_{\vec{k}} + a_{-\vec{k}}^*), \quad \psi_{\vec{k}} = -\iota\sqrt{\frac{2\omega_k}{k}}(a_{\vec{k}} - a_{-\vec{k}}^*). \tag{21}$$

In terms of $a_{\vec{k}}$ equations (16) turns to one equation

$$\frac{\partial a_{\vec{k}}}{\partial t} = -\iota\frac{\delta H}{\delta a_{\vec{k}}^*}. \tag{22}$$

Now

$$H_0 = \int \omega_k |a_{\vec{k}}|^2 d^2\vec{k}. \tag{23}$$

Then

$$H_1 = H_1^{(0,3)} + H_1^{(1,2)}. \tag{24}$$

Here

$$H_1^{(0,3)} = \frac{1}{6}\int V_{\vec{k}\vec{k}_1\vec{k}_2}^{(0,3)}(a_{\vec{k}}a_{\vec{k}_1}a_{\vec{k}_2} + a_{\vec{k}}^*a_{\vec{k}_1}^*a_{\vec{k}_2}^*)\delta(\vec{k} + \vec{k}_1 + \vec{k}_2)d\vec{k}d\vec{k}_1 d\vec{k}_2 \tag{25}$$

$$H_1^{(1,2)} = \frac{1}{2}\int V_{\vec{k}\vec{k}_1\vec{k}_2}^{(1,2)}(a_{\vec{k}}^*a_{\vec{k}_1}a_{\vec{k}_2} + a_{\vec{k}}a_{\vec{k}_1}^*a_{\vec{k}_2}^*)\delta(\vec{k} - \vec{k}_1 - \vec{k}_2)d\vec{k}d\vec{k}_1 d\vec{k}_2. \tag{26}$$

In a similar way

$$H_2 = H_2^{(0,4)} + H_2^{(1,3)} + H_2^{(2,2)}. \tag{27}$$

Only the last term in $H_2$ is important for us

$$H_2^{(2,2)} = \frac{1}{4}\int V_{\vec{k}\vec{k}_1\vec{k}_2\vec{k}_3}^{(2,2)} a_{\vec{k}}a_{\vec{k}_1}a_{\vec{k}_2}^*a_{\vec{k}_3}^*\delta(\vec{k} + \vec{k}_1 - \vec{k}_2 - \vec{k}_3)d\vec{k}d\vec{k}_1 d\vec{k}_2 d\vec{k}_3. \tag{28}$$

Explicit expressions for $V_{\vec{k}\vec{k}_1\vec{k}_2}^{(0,3)}$, $V_{\vec{k}\vec{k}_1\vec{k}_2}^{(1,2)}$, and $V_{\vec{k}\vec{k}_1\vec{k}_2\vec{k}_3}^{(2,2)}$ are presented in Appendix B.

Now everything depends on shape of function $\omega(k)$. If resonant conditions (5) have real solutions, one can neglect $H_2$ and even $H_3^{(0,3)}$. Now equations (22) takes a simple form

$$\dot{a}_{\vec{k}} + \iota\omega_k a_{\vec{k}} =$$
$$-\frac{\iota}{2}\int\{V_{\vec{k}\vec{k}_1\vec{k}_2}^{(1,2)} a_{\vec{k}_1} a_{\vec{k}_2}\delta(\vec{k} - \vec{k}_1 - \vec{k}_2) + 2V_{\vec{k}_1\vec{k}\vec{k}_2}^{(1,2)} a_{\vec{k}_1} a_{\vec{k}_2}^*\delta(\vec{k} - \vec{k}_1 + \vec{k}_2)\}d\vec{k}_1 d\vec{k}_2. \tag{29}$$

Thereafter we assume

$$V^{(1,2)}_{\vec{0},\vec{k},-\vec{k}} = 0. \tag{30}$$

For surface waves this condition is satisfied.

Equation (29) has a solution which can be treated as a stationary wave

$$a_{\vec{k}} = \sum_{n=1}^{\infty}\left[a_n e^{-\imath n\Omega t}\delta(\vec{k} - n\vec{k}_0) + b_n e^{\imath n\Omega t}\delta(\vec{k} + n\vec{k}_0)\right]. \tag{31}$$

We put $a_1 = \varepsilon$. Here $\vec{k}_0$ — an arbitrary wave vector. Coefficients $a_n$ are presented by power series

$$a_n = \varepsilon^n(a_n^{(0)} + \varepsilon a_n^{(1)} + \ldots),$$

while coefficients $b_n$ look as follows

$$b_n = \varepsilon^{n+2}(b_n^{(0)} + \varepsilon b_n^{(1)} + \ldots).$$

The frequency of the stationary wave is presented by a series in even powers of $\varepsilon$

$$\Omega = \omega(k_0) + \varepsilon^2\Delta_1 + \varepsilon^4\Delta_2 + \ldots \tag{32}$$

Now let us suppose $\varepsilon \to 0$. From (32) we see that the first nonlinear correction to frequency is proportional to $\varepsilon^2$. Now we will show that the solution (31) is unstable and the growth rate of instability is proportional to $\varepsilon$. It means that all nonlinear corrections to $a_n$, $b_n$, and $\Delta_n$ can be neglected, and one should look for a solution in the following form

$$a_{\vec{k}} = \varepsilon e^{-\imath\omega(k_0)t}\delta(\vec{k} - \vec{k}_0) + \alpha(t)e^{-\imath\omega(\kappa_1)t}\delta(\vec{k} - \vec{\kappa}_1) + \beta(t)e^{-\imath\omega(\kappa_2)t}\delta(\vec{k} - \vec{\kappa}_2), \tag{33}$$

where $\vec{\kappa}_1 + \vec{\kappa}_2 = \vec{k}_0$. Then we linearize the equation (29) and find that $\alpha$ and $\beta$ obey to the system of ordinary differential equations

$$\begin{aligned}\dot{\alpha} &= \imath e^{-\imath\Delta}\varepsilon V\beta^*, \quad V = V^{(1,2)}_{\vec{k}_0\vec{\kappa}_1\vec{\kappa}_2},\\ \dot{\beta} &= \imath e^{-\imath\Delta}\varepsilon V\alpha^*, \quad \Delta = \omega(k_0) - \omega(\kappa_1) - \omega(\kappa_2).\end{aligned} \tag{34}$$

A general solution of equation (34) is

$$\alpha = \alpha_0 e^{(-\imath\Delta/2\pm\gamma)t}, \quad \beta = \beta_0 e^{(-\imath\Delta/2\pm\gamma)t}. \tag{35}$$

Here

$$\gamma = \sqrt{\varepsilon^2|V|^2 - \frac{1}{4}\Delta^2}, \tag{36}$$

and $\alpha_0, \beta_0$ are connected by relation

$$\left(-\frac{\iota\Delta}{2} \pm \gamma\right)\beta_0 = \iota\varepsilon V \alpha_0^*. \tag{37}$$

Instability takes place if wave vectors $\kappa_1$, $\kappa_2$ are posed near the surface (or the curve)

$$\begin{aligned}\omega(k_0) &= \omega(\kappa_1) + \omega(\kappa_2), \\ \vec{k}_0 &= \vec{\kappa}_1 + \vec{\kappa}_2.\end{aligned} \tag{38}$$

The maximum of the growth rate $\gamma = \varepsilon|V|$ is reached exactly on the resonant surface, which is represented on Figure 1. The plot of growth rate (36) on a discrete grid of wave numbers for $2\pi \times 2\pi$ periodic box is given in Figure 2.

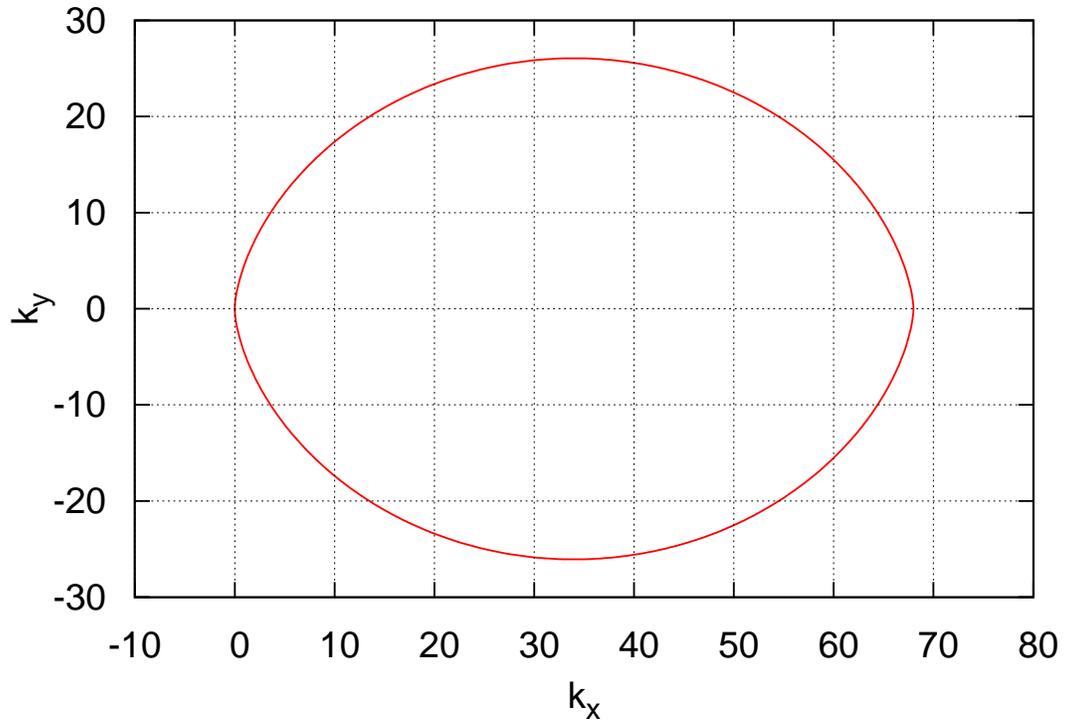

Figure 1: Resonant curve for decay of monochromatic capillary wave with $\vec{k} = (68, 0)$.

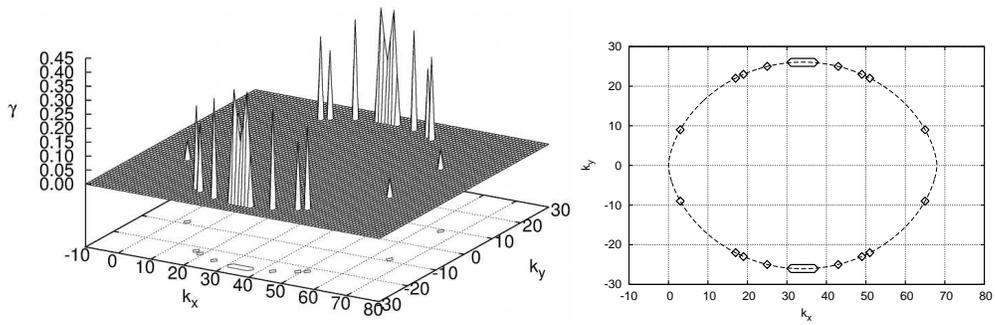

Figure 2: Growth rate for decay instability of monochromatic capillary wave with $\vec{k}_0 = (68, 0)$ and average steepness $\mu = 0.05$ on the discrete grid. Left panel: isometric projection; right panel: contour of the surface at the level $10^{-23}$.

## 3. Four-waves instability.

The theory developed in the previous chapter is applicable to study of capillary waves instability. In the case of gravity waves resonant conditions (38) have no real solutions. It means that cubic terms in the Hamiltonian can be excluded by a proper canonical transformation given by power series [22, 23]

$$a_{\vec{k}} = b_{\vec{k}}^{(0)} + b_{\vec{k}}^{(1)} + b_{\vec{k}}^{(2)} + \ldots$$
$$b_{\vec{k}}^{(0)} = a_{\vec{k}}.$$
(39)

Next terms $b_{\vec{k}}^{(1)}$, $b_{\vec{k}}^{(2)}$ are presented in Appendix B.

After canonical transformation the Hamiltonian is reduced to the form

$$H = H_0 + \tilde{H}_2, \quad (40)$$

$$\tilde{H}_2 = \frac{1}{4} \int T_{\vec{k}\vec{k}_1\vec{k}_2\vec{k}_3} b^*_{\vec{k}} b^*_{\vec{k}_1} b_{\vec{k}_2} b_{\vec{k}_3} \delta(\vec{k} + \vec{k}_1 - \vec{k}_2 - \vec{k}_3) d\vec{k} d\vec{k}_1 d\vec{k}_2 d\vec{k}_3. \quad (41)$$

Explicit and complicated expression for $T_{\vec{k}\vec{k}_1\vec{k}_2\vec{k}_3}$ is given in Appendix B. $T_{\vec{k}\vec{k}_1\vec{k}_2\vec{k}_3}$ is a homogeneous function of order three.

Now canonical variable $b_{\vec{k}}$ obeys so called "Zakharov's equation" [10]

$$\dot{b}_{\vec{k}} + \imath \omega_k b_{\vec{k}} = -\frac{\imath}{2} \int T_{\vec{k}\vec{k}_1\vec{k}_2\vec{k}_3} b^*_{\vec{k}_1} b_{\vec{k}_2} b_{\vec{k}_3} \delta(\vec{k} + \vec{k}_1 - \vec{k}_2 - \vec{k}_3) d\vec{k}_1 d\vec{k}_2 d\vec{k}_3. \quad (42)$$

Equation (42) has exact solution

$$b_{\vec{k}} = A \delta(\vec{k} - \vec{k}_0) e^{-\imath \Omega_0 t},$$
$$\Omega_0 = \omega(k_0) + \frac{1}{2} T_{\vec{k}_0} |A|^2, \quad (43)$$
$$T_{\vec{k}_0} = T_{\vec{k}_0 \vec{k}_0 \vec{k}_0 \vec{k}_0} = \frac{1}{2\pi} k_0^3.$$

This solution is nothing but the stationary Stokes wave. It gives right expression for two first terms in series (1) and (2).

Equation (42) has also reach set of approximate quasi-periodic solutions. Let $\vec{\kappa}_1$ and $\vec{\kappa}_2$ are two arbitrary wave vectors. In the limit of small $|A_1|^2$, $|A_2|^2$ equation has the following solution:

$$b_{\vec{k}} = A_1 \delta(\vec{k} - \vec{\kappa}_1) e^{-\imath \tilde{\Omega}_1 t} + A_2 \delta(\vec{k} - \vec{\kappa}_2) e^{-\imath \tilde{\Omega}_2 t} + \ldots \quad (44)$$

$$\tilde{\Omega}_1 = \omega(\kappa_1) + \frac{1}{2} T_{\vec{\kappa}_1} |A_1|^2 + T_{\vec{\kappa}_1, \vec{\kappa}_2} |A_2|^2,$$
$$\tilde{\Omega}_2 = \omega(\kappa_2) + T_{\vec{\kappa}_1, \vec{\kappa}_2} |A_1|^2 + \frac{1}{2} T_{\vec{\kappa}_1} |A_2|^2.$$
(45)

Here $T_{\vec{\kappa}_1,\vec{\kappa}_2} = T_{\vec{\kappa}_1\vec{\kappa}_2,\vec{\kappa}_1\vec{\kappa}_2}$. Equations (44), (45) are valid if nonlinear terms in (45) are much less that linear.

In the particular case $\vec{\kappa}_2 = -\vec{\kappa}_1$, $|A_2| = |A_1|$ solution (44) is just a standing wave.

Both propagating wave (43) and standing wave (44) are unstable. To study instability of propagating wave (43) we will look for a solution in the following form

$$b_{\vec{k}} = A_0 \delta(\vec{k} - \vec{k}_0) e^{-\imath \Omega_0 t} + \alpha \delta(\vec{k} - \vec{k}_0 - \vec{\kappa}) e^{-\imath \Omega_1 t} + \beta \delta(\vec{k} - \vec{k}_0 + \vec{\kappa}) e^{-\imath \Omega_2 t}. \quad (46)$$

Here

$$\begin{aligned}
\Omega_1 &= \omega(\vec{k}_0 + \vec{\kappa}) + 2T(\vec{k}_0, \vec{k}_0 + \vec{\kappa})|A|^2, \\
\Omega_2 &= \omega(\vec{k}_0 - \vec{\kappa}) + 2T(\vec{k}_0, \vec{k}_0 - \vec{\kappa})|A|^2.
\end{aligned} \quad (47)$$

By plugging (44) into (42) and linearizing over $\alpha$ and $\beta$ we set system of ordinary differential equations, similar to (34)

$$\begin{aligned}
\dot{\alpha} &= \tfrac{\imath}{2} e^{-\imath \Delta} |A_0|^2 T \beta^*, \quad T = T_{\vec{k}_0,\vec{k}_0,\vec{k}_0+\vec{\kappa},\vec{k}_0-\vec{\kappa}}, \\
\dot{\beta} &= \tfrac{\imath}{2} e^{-\imath \Delta} |A_0|^2 T \alpha^*, \\
\Delta &= 2\omega(k_0) - \omega(\vec{k}_0 - \vec{\kappa}) - \omega(\vec{k}_0 + \vec{\kappa}) + 2(T_{\vec{k}_0} - T_{\vec{k}_0,\vec{\kappa}_1} - T_{\vec{k}_0,\vec{\kappa}_2})|A_0|^2.
\end{aligned} \quad (48)$$

Solutions of equation (48) are given by formulae (35) where

$$\gamma = \sqrt{|A|^4 |T|^2 - \frac{1}{4}\Delta^2}. \quad (49)$$

One can see that instability occurs in a vicinity of the curve $\Delta = 0$. For waves on a deep water we can put $\vec{k}_0 = \vec{i}$, $\vec{\kappa} = x\vec{i} + y\vec{j}$. Then the condition $\Delta = 0$ is reduced to the famous Phillips curve [24]

$$[(1 + x)^2 + y^2]^{1/4} + [(1 - x)^2 + y^2]^{1/4} = 2. \quad (50)$$

Here $-5/4 \leq x \leq 5/4$. The Phillips curve is plotted in Figure 3. The coupling coefficient $T(\vec{k}_0, x)$, evaluated on the Phillips curve, can be presented in the form

$$T(2\vec{k}_0, \vec{k}_0 + \vec{\kappa}, \vec{k}_0 - \vec{\kappa}) = k_0^3 f(x), \quad x = \frac{\kappa_x}{k_0},$$

$f(-x) = f(x)$ symmetric function. It is important to mention that $f(5/4) = 0$. This fact was first discovered by Dyachenko and Zakharov in 1994 [25]. Decreasing of $f(x)$ with growth of $x$ means that the four-wave instability is mostly

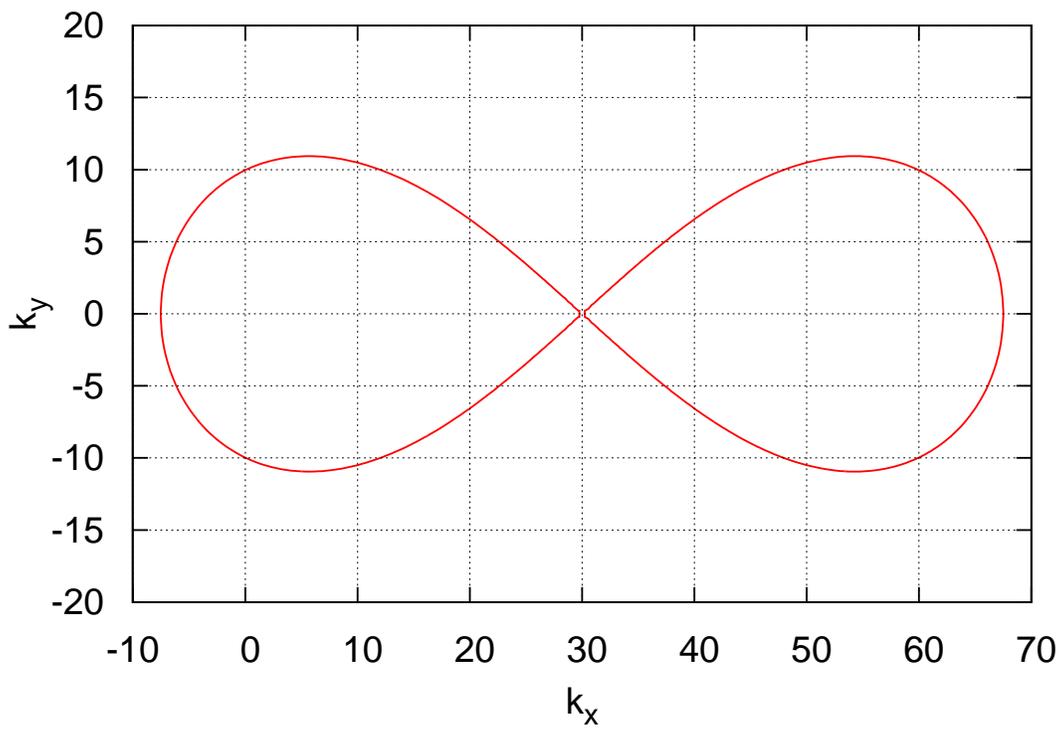

Figure 3: Phillips curve for $\vec{k}_0 = (30, 0)$.

modulational, because the most unstable modes are concentrated at $\kappa \to 0$. In this region

$$\gamma \simeq \frac{1}{2} \sqrt{-2T|A|^2 \frac{\partial^2 \omega}{\partial k_\alpha \partial k_\beta} \kappa_\alpha \kappa_\beta - \left(\frac{\partial^2 \omega}{\partial k_\alpha \partial k_\beta} \kappa_\alpha \kappa_\beta\right)^2}. \tag{51}$$

Instability is concentrated inside the angle where

$$\frac{\partial^2 \omega}{\partial k_\alpha \partial k_\beta} \kappa_\alpha \kappa_\beta < 0.$$

The plot of growth rate (49) on a discrete grid of wave numbers for $2\pi \times 2\pi$ periodic box is given in Figure 4. Four-wave instability of propagating Stokes

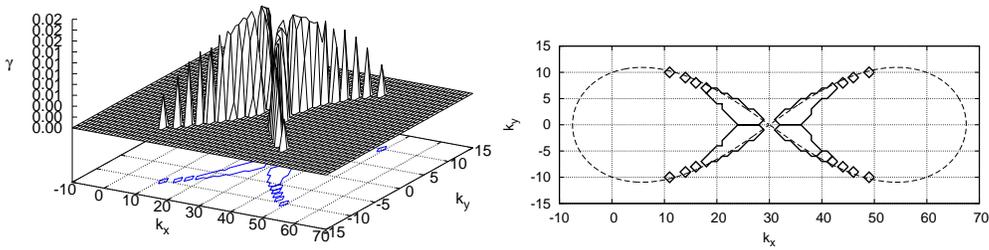

Figure 4: Growth rate for Phillips instability of monochromatic gravity wave with $\vec{k}_0 = (30, 0)$ and average steepness $\mu = 0.1$ on the discrete grid. Left panel: isometric projection; right panel: contour of the surface at the level $10^{-23}$.

waves was studied in details in many papers (e. g. [6, 7, 8, 9, 10, 11, 26]).

Let us study instability of by-harmonic wave (46). We concentrate only on the case of standing wave $\kappa_2 = -\kappa_1 = k_0$. The standing wave is unstable due to a different mechanism. Firstly, each propagating wave composing the standing wave endure its own modulational instability. Then another instability appears. In this new type of instability we have simultaneous excitation of two waves with wave numbers $\pm\vec{\kappa}$ such that $|\kappa| = |k_0|$. The maximal growth-rate of this instability

$$\gamma_{max} \simeq \frac{1}{2} \tilde{T} |A|^2.$$

Here

$$\tilde{T} = T_{\vec{k}_0, -\vec{k}_0, \vec{\kappa}, -\vec{\kappa}} = k_0^3 f(\cos\theta), \tag{52}$$

and $\theta$ is an angle between $\vec{k}_0$ and $\vec{\kappa}$.

## 4. Numerical Simulation Scheme.

The problem of numerical integration of system of equations (19) is rather hard. One of the most important questions is what time integration algorithm to choose. According to the property of the equations it would be natural to develop a numerical integration scheme conserving the Hamiltonian. Let us follow the article [27]. One can introduce a discrete variation of Hamiltonian (17) on one step on time $H^n \longrightarrow H^{n+1}$

$$\Delta H = H^{n+1} - H^n. \tag{53}$$

The Hamiltonian is a function of canonical variables $\eta$ and $\psi$. The discrete variations of these functions on a time step are equal to

$$\Delta \eta = \eta^{n+1} - \eta^n, \quad \Delta \psi = \psi^{n+1} - \psi^n. \tag{54}$$

One can expand discrete variation $\Delta H$ via $\Delta \eta$ and $\Delta \psi$ (that is done in Appendix A)

$$\Delta H = H_\psi \Delta \psi + H_\eta \Delta \eta. \tag{55}$$

It is easy to see that $H_\eta$ and $H_\psi$ are discrete analogues to the continuous variations $\frac{\delta H}{\delta \eta}$ and $\frac{\delta H}{\delta \psi}$.

One can demand conservation of Hamiltonian $\Delta H / \tau = 0$ during time step $\tau$. Obviously, this equality can take place if the following conditions are valid

$$\begin{aligned}\frac{\Delta \eta}{\tau} &= H_\psi, \\ \frac{\Delta \psi}{\tau} &= -H_\eta.\end{aligned} \tag{56}$$

In some sense this is a discrete analogue of Hamiltonian equations (16). Thus, if Hamiltonian variation (53) is expanded via variations $\Delta \eta$ and $\Delta \psi$, it is possible to get equations (56).

As it was mentioned above it is more convenient to rewrite the equations in terms of Fourier harmonics. Using the results obtained in Appendix A (A.2-A.7) one can obtain an implicit difference scheme

$$\begin{aligned}\frac{\eta_{\vec{k}}^{n+1} - \eta_{\vec{k}}^n}{\tau} =\ & \tfrac{1}{2}|\vec{k}|\left(\psi_{\vec{k}}^{n+1} + \psi_{\vec{k}}^n\right) - \\ & -\tfrac{1}{4}\hat{F}\left(\nabla, (\eta^{n+1} + \eta^n)\nabla(\psi^{n+1} + \psi^n)\right) - \\ & -\tfrac{1}{4}|\vec{k}|\hat{F}\left((\eta^{n+1} + \eta^n)\hat{k}(\psi^{n+1} + \psi^n)\right) + \\ & +\tfrac{1}{4}|\vec{k}|\hat{F}\left[\left(\eta^{n+1} + \eta^n\right)\hat{k}\left(\eta^{n+1}\hat{k}\psi^{n+1} + \eta^n\hat{k}\psi^n\right)\right] - \\ & -\tfrac{1}{8}|\vec{k}|^2\hat{F}\left[((\eta^{n+1})^2 + (\eta^n)^2)\hat{k}(\psi^{n+1} + \psi^n)\right] + \\ & +\tfrac{1}{8}|\vec{k}|\hat{F}\left[((\eta^{n+1})^2 + (\eta^n)^2)\nabla^2(\psi^{n+1} + \psi^n)\right].\end{aligned} \tag{57}$$

$$\frac{\psi_{\vec{k}}^{n+1} - \psi_{\vec{k}}^{n}}{\tau} = -\frac{1}{2}\frac{\omega_k^2}{|\vec{k}|}\left(\eta_{\vec{k}}^{n+1} + \eta_{\vec{k}}^{n}\right) - \\
-\frac{1}{4}\hat{F}\left(|\nabla\psi^{n+1}|^2 + |\nabla\psi^n|^2\right) + \\
+\frac{1}{4}\hat{F}\left((\hat{k}\psi^{n+1})^2 + (\hat{k}\psi^n)^2\right) - \\
-\frac{1}{4}\hat{F}\left[\hat{k}\left(\psi^{n+1} + \psi^n\right)\hat{k}\left(\eta^{n+1}\hat{k}\psi^{n+1} + \eta^n\hat{k}\psi^n\right)\right] - \\
-\frac{1}{4}\hat{F}\left[(\eta^{n+1} + \eta^n)(\nabla^2\psi^{n+1}\hat{k}\psi^{n+1} + \nabla^2\psi^n\hat{k}\psi^n)\right]. \quad (58)$$

Here $\hat{F}$ is the Fourier transform operator.

It is useful to resolve linear part of scheme (57-58) with respect to $\eta^{n+1}$ and $\psi^{n+1}$. Let us denote nonlinear terms in right hand sides of these equations as:

$$\begin{aligned}
R_\eta^{n+1} &= -\frac{1}{4}\hat{F}\left(\nabla, (\eta^{n+1} + \eta^n)\nabla(\psi^{n+1} + \psi^n)\right) - \\
&\quad -\frac{1}{4}|\vec{k}|\hat{F}\left((\eta^{n+1} + \eta^n)\hat{k}(\psi^{n+1} + \psi^n)\right) + \\
&\quad +\frac{1}{4}|\vec{k}|\hat{F}\left[(\eta^{n+1} + \eta^n)\hat{k}\left(\eta^{n+1}\hat{k}\psi^{n+1} + \eta^n\hat{k}\psi^n\right)\right] - \\
&\quad -\frac{1}{8}|\vec{k}|^2\hat{F}\left[((\eta^{n+1})^2 + (\eta^n)^2)\hat{k}(\psi^{n+1} + \psi^n)\right] + \\
&\quad +\frac{1}{8}|\vec{k}|\hat{F}\left[((\eta^{n+1})^2 + (\eta^n)^2)\nabla^2(\psi^{n+1} + \psi^n)\right], \\
R_\psi^{n+1} &= -\frac{1}{4}\hat{F}\left(|\nabla\psi^{n+1}|^2 + |\nabla\psi^n|^2\right) + \\
&\quad +\frac{1}{4}\hat{F}\left((\hat{k}\psi^{n+1})^2 + (\hat{k}\psi^n)^2\right) - \\
&\quad -\frac{1}{4}\hat{F}\left[\hat{k}\left(\psi^{n+1} + \psi^n\right)\hat{k}\left(\eta^{n+1}\hat{k}\psi^{n+1} + \eta^n\hat{k}\psi^n\right)\right] - \\
&\quad -\frac{1}{4}\hat{F}\left[(\eta^{n+1} + \eta^n)(\nabla^2\psi^{n+1}\hat{k}\psi^{n+1} + \nabla^2\psi^n\hat{k}\psi^n)\right].
\end{aligned} \quad (59)$$

Using these notations discrete scheme can be written as follows

$$\begin{aligned}
\eta_{\vec{k}}^{n+1} &= A(k,\tau)\eta_{\vec{k}}^n + B(k,\tau)\psi_{\vec{k}}^n + C(k,\tau)R_\eta^{n+1} + D(k,\tau)R_\psi^{n+1}, \\
\psi_{\vec{k}}^{n+1} &= E(k,\tau)\eta_{\vec{k}}^n + A(k,\tau)\psi_{\vec{k}}^n + F(k,\tau)R_\eta^{n+1} + C(k,\tau)R_\psi^{n+1}.
\end{aligned} \quad (60)$$

Here

$$A(k,\tau) = \frac{1 - \frac{1}{4}\omega_k^2\tau^2}{1 + \frac{1}{4}\omega_k^2\tau^2}, \quad B(k,\tau) = \frac{\tau k}{1 + \frac{1}{4}\omega_k^2\tau^2},$$

$$C(k,\tau) = \frac{\tau}{1 + \frac{1}{4}\omega_k^2\tau^2}, \quad D(k,\tau) = \frac{1}{2}\tau B(k,\tau), \quad (61)$$

$$E(k,\tau) = -\frac{\omega_k^2}{k}C(k,\tau), \quad F(k,\tau) = \frac{1}{2}\tau E(k,\tau).$$

Thus, we get implicit (terms $R_\eta^{n+1}$ and $R_\psi^{n+1}$ contain $\eta_{\vec{k}}^{n+1}$ and $\psi_{\vec{k}}^{n+1}$) difference scheme. The important feature of this scheme is that conservation of Hamiltonian (13)-(14) is embedded in it.

The implicit numerical scheme (60) can be solved by method of fixed-point iterations [28]. Let us write this procedure for $\eta_{\vec{k}}^{n+1,s}$ and $\psi_{\vec{k}}^{n+1,s}$, here $s$ is an iteration number. Corresponding to (60) one can get

$$\begin{aligned}
\eta_{\vec{k}}^{n+1,s+1} &= A(k,\tau)\eta_{\vec{k}}^n + B(k,\tau)\psi_{\vec{k}}^n + C(k,\tau)R_\eta^{n+1,s} + D(k,\tau)R_\psi^{n+1,s}, \\
\psi_{\vec{k}}^{n+1,s+1} &= E(k,\tau)\eta_{\vec{k}}^n + A(k,\tau)\psi_{\vec{k}}^n + F(k,\tau)R_\eta^{n+1,s} + C(k,\tau)R_\psi^{n+1,s}; \quad (62) \\
\eta_{\vec{k}}^{n+1,0} &= \eta_{\vec{k}}^n, \; \psi_{\vec{k}}^{n+1,0} = \psi_{\vec{k}}^n.
\end{aligned}$$

Iterations continue until the desired accuracy of Hamiltonian conservation $\epsilon$ is achieved. Convergence is ensured by the choice of time step $\tau$ to be small enough. Sufficient condition for convergence of iterations is the following:

$$\tau^2 < \frac{1}{\max[(\vec{\nabla}\psi)^2 + (\hat{k}\psi)^2]k^2 - 1/4\min(\omega_{\vec{k}}^2)}. \qquad (63)$$

Taking into account that the first factor in the first term in the denominator of (63) is of the order of maximum velocity on the surface, this condition is similar to $\tau < \Delta x/v_{max}$. We resolved linear part of the equation this is why frequency doesn't come into play.

In most cases it is enough to follow the convergence of the relative error

$$\frac{\sum_{\vec{k}} \left|\eta_{\vec{k}}^{n+1,s+1}\right|^2 - \sum_{\vec{k}} \left|\eta_{\vec{k}}^{n+1,s}\right|^2}{\sum_{\vec{k}} \left|\eta_{\vec{k}}^{n+1,s}\right|^2} < \epsilon. \qquad (64)$$

When studying gravity waves this condition is equivalent to calculation of potential energy with desired accuracy. For the weakly nonlinear regime quadratic part of Hamiltonian is dominant, so the physical meaning of this condition is quite clear.

The condition on time step (63) is not very practical to use. It is much easier to control step on time by demand of iteration process convergence to the desired accuracy in a number of iterations less than some given $N_{max}$ and more than some given $N_{min}$.

This numerical scheme can be used for simulation of freely decaying waves. For simulation of turbulence we should introduce pumping and damping terms in the equations. It is possible to do that by different ways. We have applied split-step method, which is very popular in numerical simulation of pulse propagation in optic fibers. Let us consider a simple example.

Let us suppose a linear damping with rate $\gamma_k$ in our model

$$\dot{\psi}_{\vec{k}} = R.H.S. - \gamma_k \psi_{\vec{k}}. \qquad (65)$$

It is possible to take into account this damping without significant changes in calculations scheme. First one can obtain solution of equations (19) without damping using iteration scheme described above. Let us denote this solution by $\tilde{\psi}_{\vec{k}}^{n+1}$. Second, the solution of the whole system of equations can be calculated by the next step

$$\psi_{\vec{k}}^{n+1} = \tilde{\psi}_{\vec{k}}^{n+1} \exp(-\gamma_k \tau). \qquad (66)$$

It is worth to say that for weak turbulence simulation the most interesting part of spectrum is in the "inertial interval" where there are no damping or pumping at all. Even more, the nature of damping and pumping is not important. In this case influence of non-conservative terms can be described by such a rough scheme. As a bonus we have eliminated the restrictions on time step

$$\max(|\gamma_k|)\tau < 1, \qquad (67)$$

unavoidable in the case of integration by standard Runge-Kutta methods.

Pumping can be considered in a similar way.

## 5. Capillary waves

In this section we briefly review our previous results published in [13] and report new observations. System of equations (19) was simulated in the domain $L_x = L_y = 2\pi$. Surface tension coefficient $\sigma = 1$. Number of grid points was $512 \times 512$. A monochromatic wave of amplitude $|a_{\vec{k}_0}| = 2 \times 10^{-3}$, which corresponds to average steepness $\mu = 0.05$, was taken as initial conditions. Its wave number vector $\vec{k}_0 = (68, 0)$. All other harmonics were of amplitude $|a_{\vec{k}}| \sim 10^{-12}$ and with random phase (Figure 5). As was mentioned above resonant curve almost never passes through grid points (there are two non trivial points $\vec{k} = (0; 0)$ and initial wave $\vec{k} = (k_0, 0)$; this process has a zero growth rate). Detailed picture of resonant curve on the grid in region with highest grid points density in the vicinity of the curve is shown in Figure 6. One can see, that some points are closer to resonant curve than others.

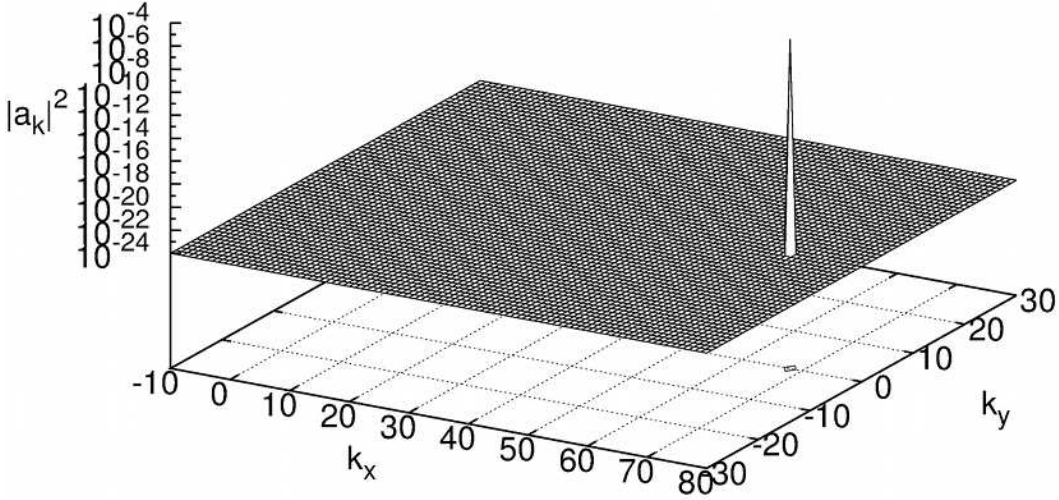

Figure 5: Decay of the monochromatic capillary wave. Initial conditions. Time $t = 0$.

In the beginning one can see growth of several harmonics as it is predicted in (35) and (36). Different stages of the decay process are represented in Figures 7-10. Time is given in periods of initial wave $T_0$. We represent isometric projection of the $|a_{\vec{k}}|^2$-surface and contour of this surface at the level $10^{-23}$ (order of magnitude higher than background noise). The full picture of the $k$-plane in the final moment of simulations is represented in Figure 11. Zoom of the most interesting region of the Figure 11 (initial decay region) is represented in Figure 12. One can see that although the amplitudes of the waves are stochastic, the spectrum is still strongly anisotropic.

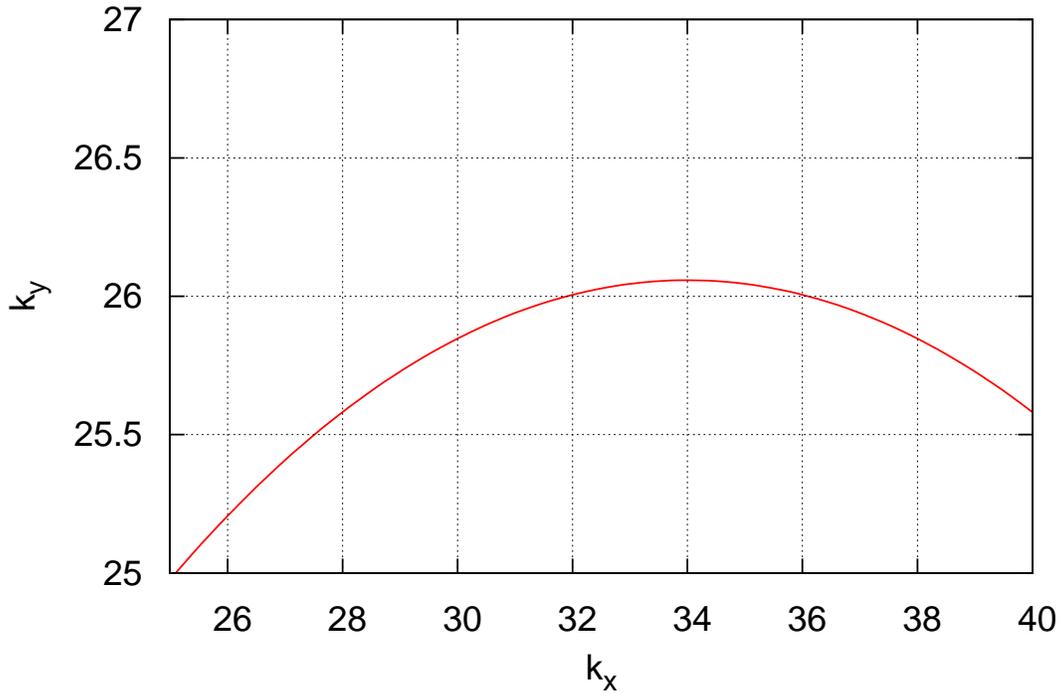

Figure 6: Part of resonant curve for decay of monochromatic capillary wave with $\vec{k}_0 = (68, 0)$. The different mismatches for different grid knots are clearly seen.

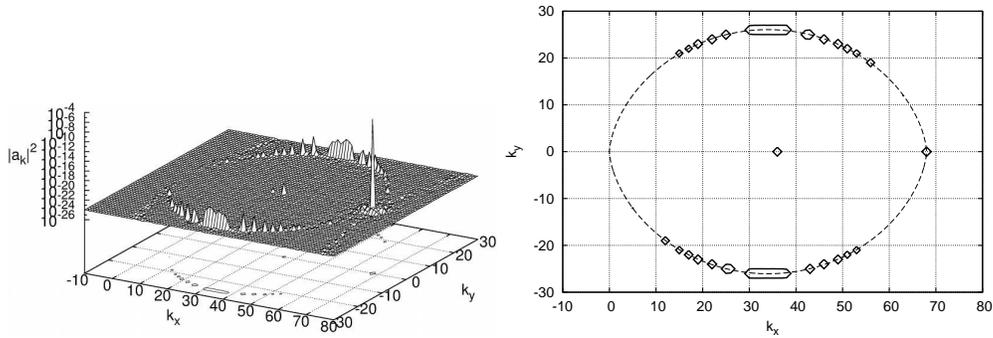

Figure 7: Decay of the monochromatic capillary wave. Growth of the harmonics in the vicinity of the resonant curve has began. Time $t = 318T0$.

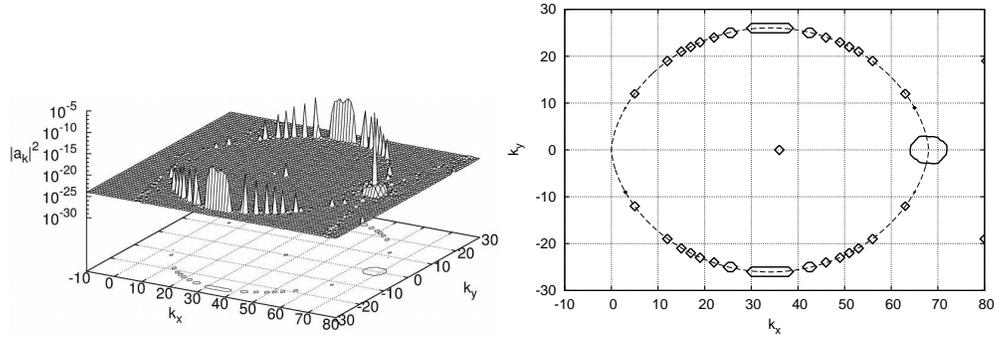

Figure 8: Decay of the monochromatic capillary wave. Decay harmonics are well developed. Time $t = 794T0$.

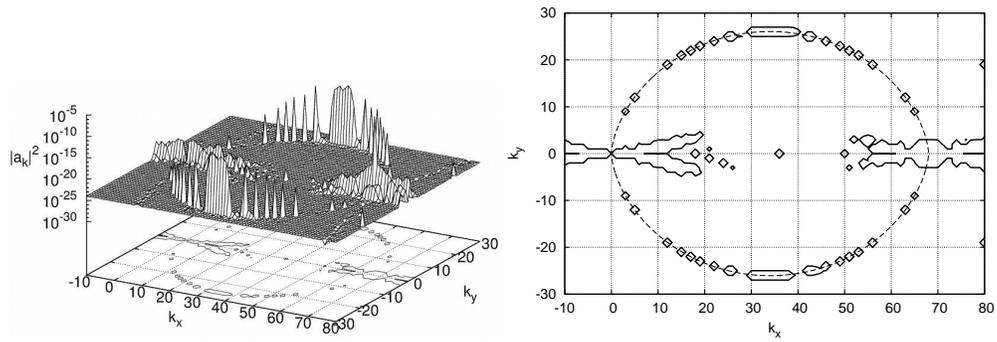

Figure 9: Decay of the monochromatic capillary wave. Secondary nonlinear processes are revealed. Time $t = 1112T0$.

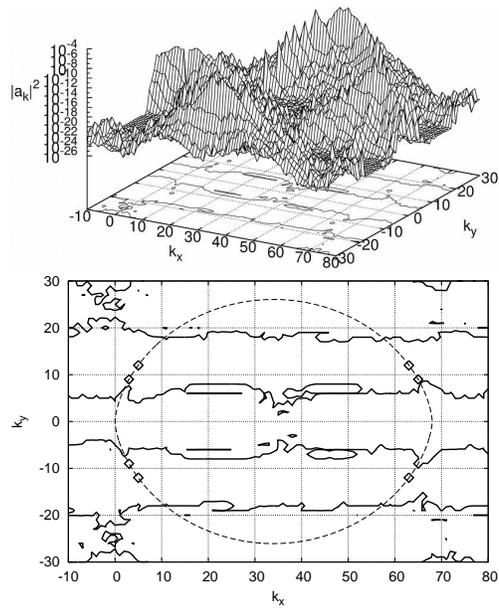

Figure 10: Decay of the monochromatic capillary wave. Secondary nonlinear processes are well developed. Time $t = 1589T0$.

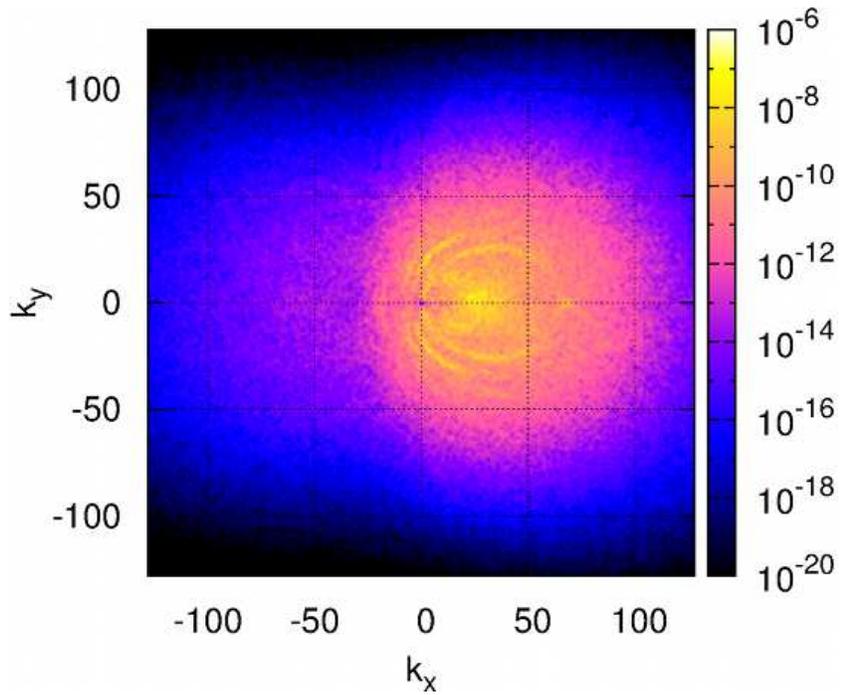

Figure 11: Decay of the monochromatic capillary wave. Full $k$-plane. Time $t = 144488T0$.

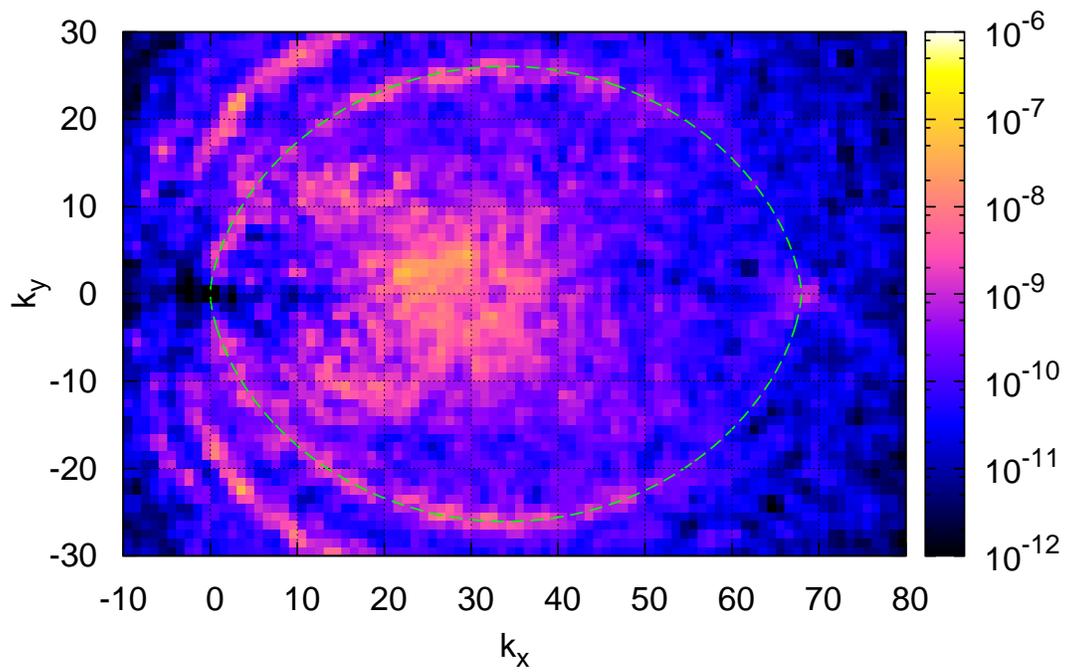

Figure 12: Decay of the monochromatic capillary wave. Zoom of the initial decay region. Time $t = 144488 T0$.

## 6. Gravity waves.

In the case of gravity waves on the surface of deep fluid the dispersion is the following

$$\omega_k = \sqrt{gk}, \tag{68}$$

here $g$ is a gravity acceleration. Here and further let us suppose $g = 1$.

In this case dispersion is of non-decay type conditions (38) have no real non-trivial solutions, and main process is four-wave scattering. Therefore one can make a substitution to eliminate third order terms corresponding to the decay process. This is the reason why we have to use Hamiltonian expansion up to forth order in the case of gravity waves.

Let us consider the same initial conditions as in the case of monochromatic capillary wave decay, i.e. one monochromatic wave and random phase noise of small amplitude. The main processes involve large amplitude of initial wave most times. In this case one wave to three and invert processes are much weaker than scattering two waves with the same amplitude and the same wave vector to two other waves.

Resonance conditions for such process are as follows

$$\omega_{k_1} + \omega_{k_2} = 2\omega_{k_0}, \quad \vec{k}_1 + \vec{k}_2 = 2\vec{k}_0. \tag{69}$$

The resonant curve for this conditions is shown in Figure 3.

System of equations (19) was simulated in domain $L_x = L_y = 2\pi$, gravity acceleration $g = 1$. Grid size was equal to $512 \times 512$ points. As an initial conditions monochromatic wave of amplitude $|a_{\vec{k}_0}| = 1.3 \times 10^{-3}$ with wave number vector $\vec{k}_0 = (30, 0)$ was used. All other harmonics were of amplitude $|a_{\vec{k}}| \sim 10^{-12}$ and random phase (Figure 13). In the beginning one can observe exponential growth of several harmonics in the vicinity of resonant curve (detailed picture of resonant curve in the surroundings of the initial wave is shown in Figure 14). This is shown on Figure 15. It is clearly seen that wave with wavevector $(33, 2)$ has smallest mismatch and, due to that, grows. As we already know four-wave scattering growth rate has highest values in the vicinity of $\vec{k}_0 = (k_0, 0)$. Due to this the initial growth is lumped about the cross of resonant manifold. Different stages are represented in Figures 16-21. We represent isometric projection of the $|a_{\vec{k}}|^2$-surface and contour of this surface at the level $10^{-23}$ (order of magnitude higher than background noise). The full picture of the $k$-plane in the final moment of simulations is represented in Figure 22. Zoom of the most interesting region of the Figure 22 (initial instability region) is represented in Figure 23. One can observe still weak but observable downshift of the spectrum.

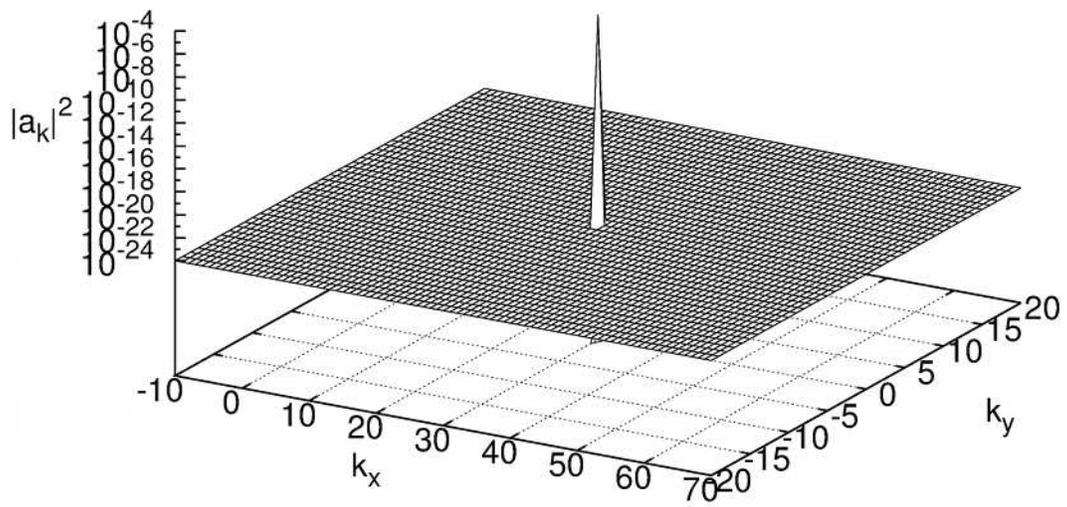

Figure 13: Instability of the monochromatic gravity wave. Initial conditions. Time $t = 0$.

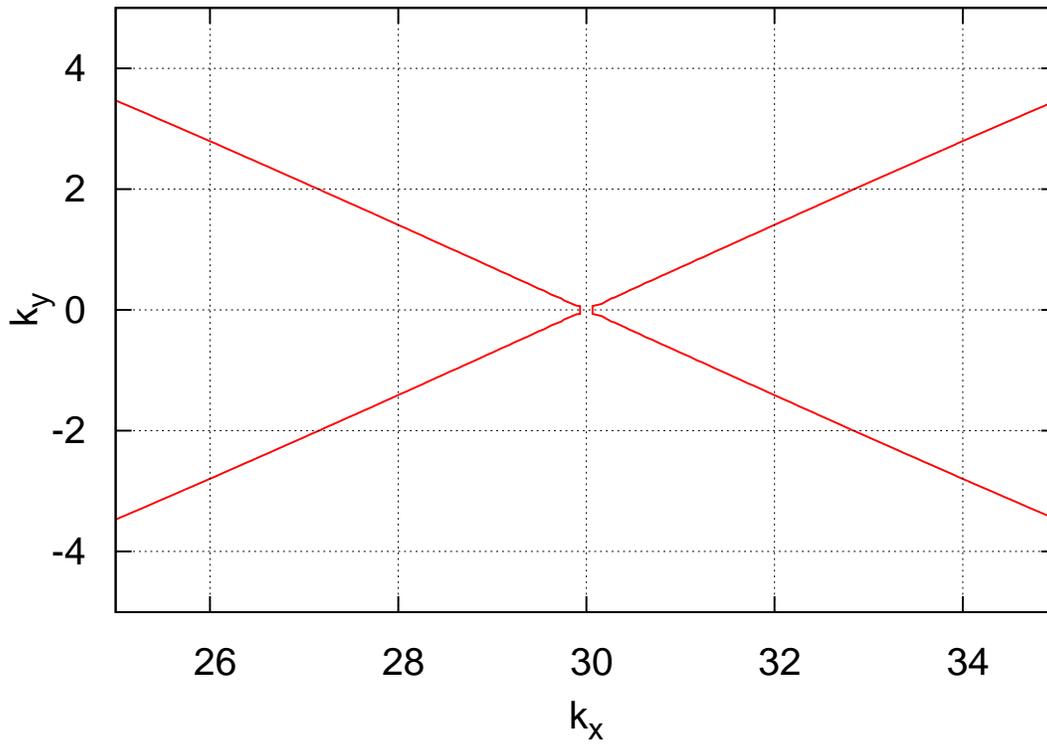

Figure 14: Gravity waves. Part of resonant curve. Different mismatch for different grid points is well seen.

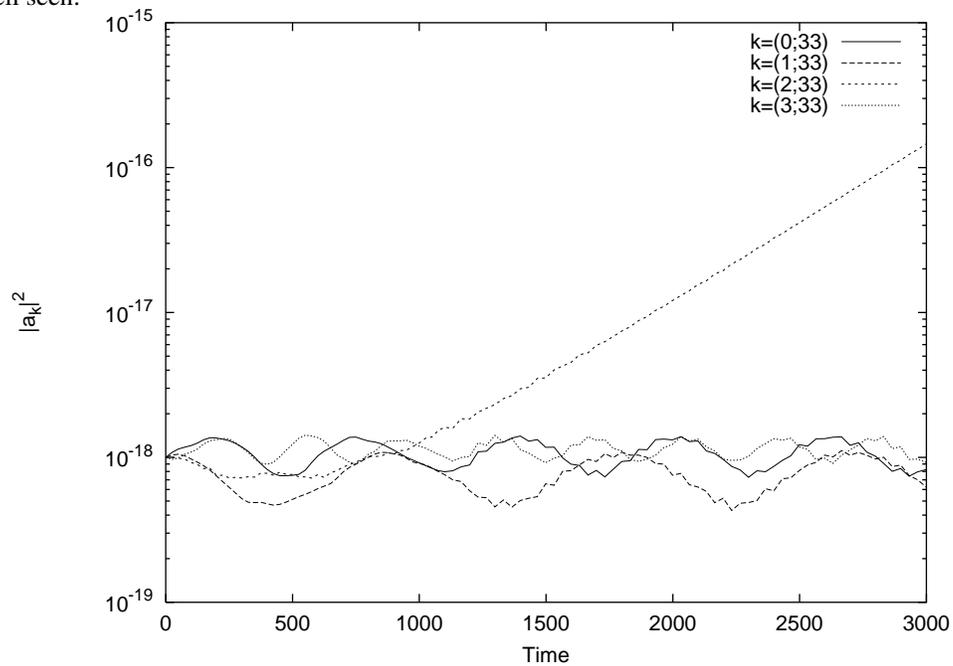
26...

Figure 15: Gravity waves. Growth of harmonics amplitude as a function on time. One can see, that harmonic $\vec{k} = (33, 2)$ is in almost exact resonance and the others are not.

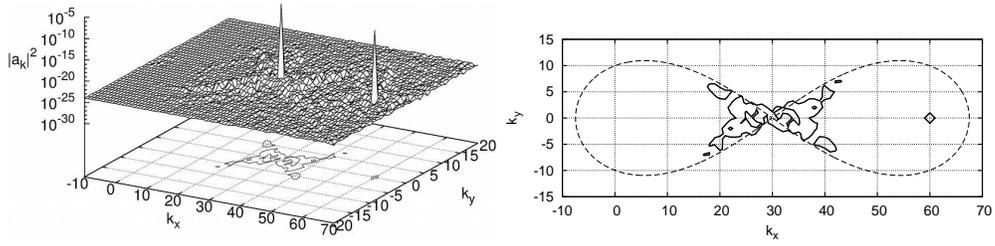

Figure 16: Instability of the monochromatic gravity wave. Growth of the harmonics in the vicinity of the resonant curve has began. Time $t = 43T0$.

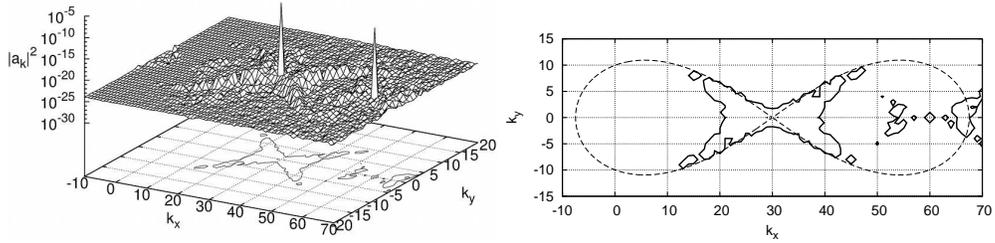

Figure 17: Instability of the monochromatic gravity wave. Growth of the harmonics in the vicinity of the resonant curve continues. Time $t = 87T0$.

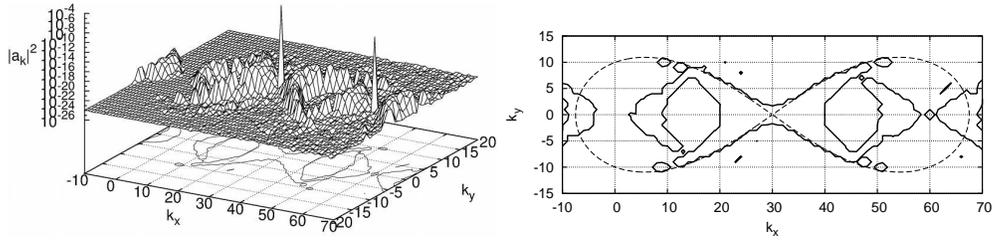

Figure 18: Instability of the monochromatic gravity wave. Harmonics on the resonant curve are well developed. Time $t = 174T0$.

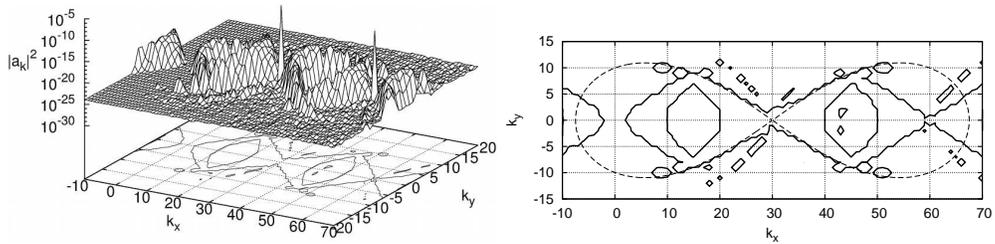

Figure 19: Instability of the monochromatic gravity wave. Beginning of the secondary processes. Time $t = 261T0$.

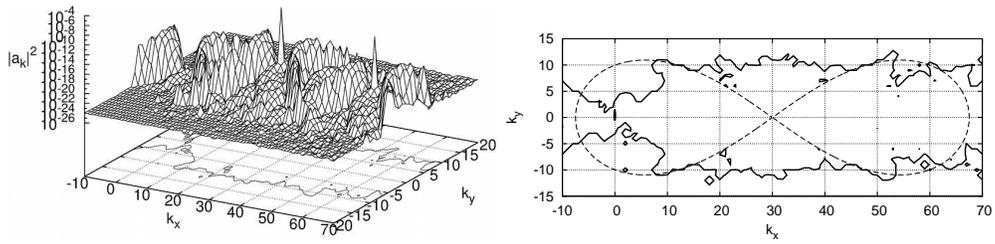

Figure 20: Instability of the monochromatic gravity wave. Secondary processes are well developed. Time $t = 348T0$.

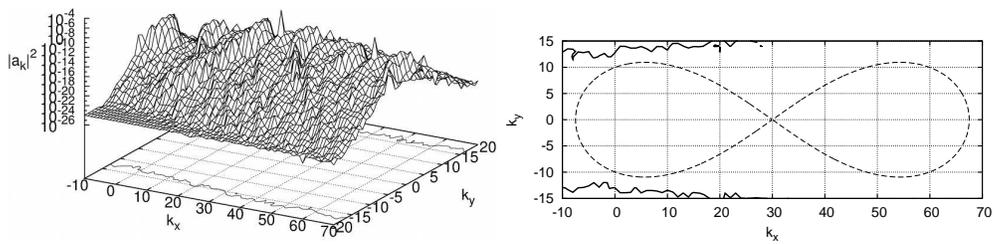

Figure 21: Instability of the monochromatic gravity wave. Secondary processes hide the structure of the resonances. Time $t = 435T0$.

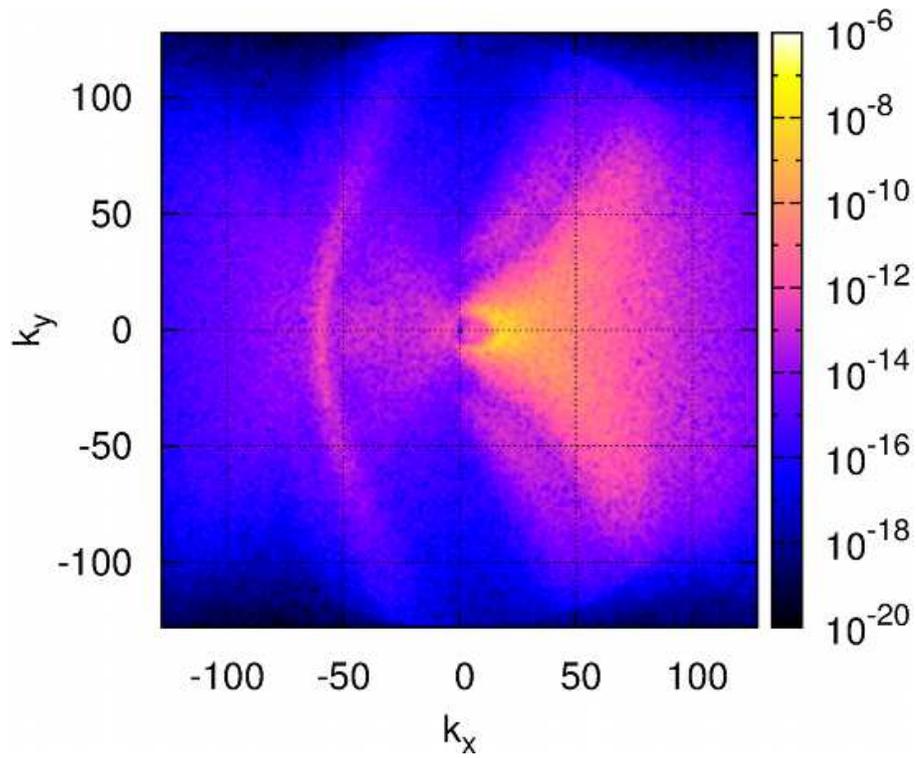

Figure 22: Instability of the monochromatic gravity wave. Full $k$-plane. Time $t = 1204T0$.

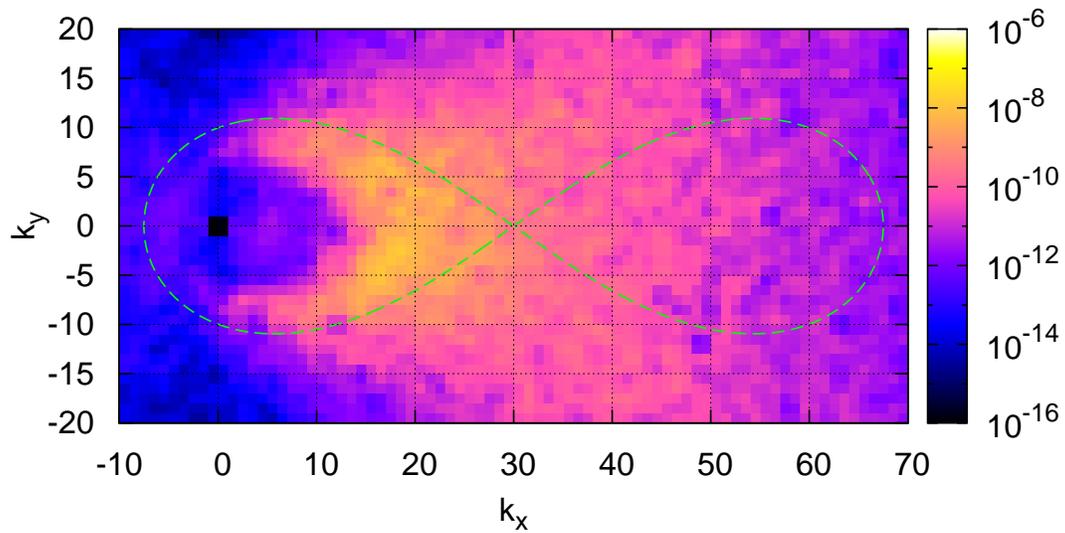

Figure 23: Decay of the monochromatic capillary wave. Zoom of the initial instability region. Time $t = 1204T0$.

## 7. Instability of the standing wave.

May be the most practically important case of surface waves instability is the case of instability of the standing wave, when we have interaction of two waves $a_{\vec{k}_0}$ and $a_{-\vec{k}_0}$. In this case resonant curve is a circle with the center at zero wave number vector and of radius $|\vec{k}_0|$. It is clear that such a process is general for any isotropic dispersion. The theory for very similar instability in plasma was developed in [29].

### 7.1. Standing capillary wave.

Simulation results for the standing capillary ($\mu = 0.1$) wave are represented in Figures 25-29. Contour plots correspond to the level $|a_{\vec{k}}|^2 = 10^{-23}$, which is order of magnitude higher than background noise. The full picture of the $k$- plane in the final moment of simulations is represented in Figure 30. Zoom of the most interesting region of the Figure 30 (initial instability region) is represented in Figure 31. We observe isotropization of the wavefield, but in order to obtain a smooth spectrum we need to wait much longer.

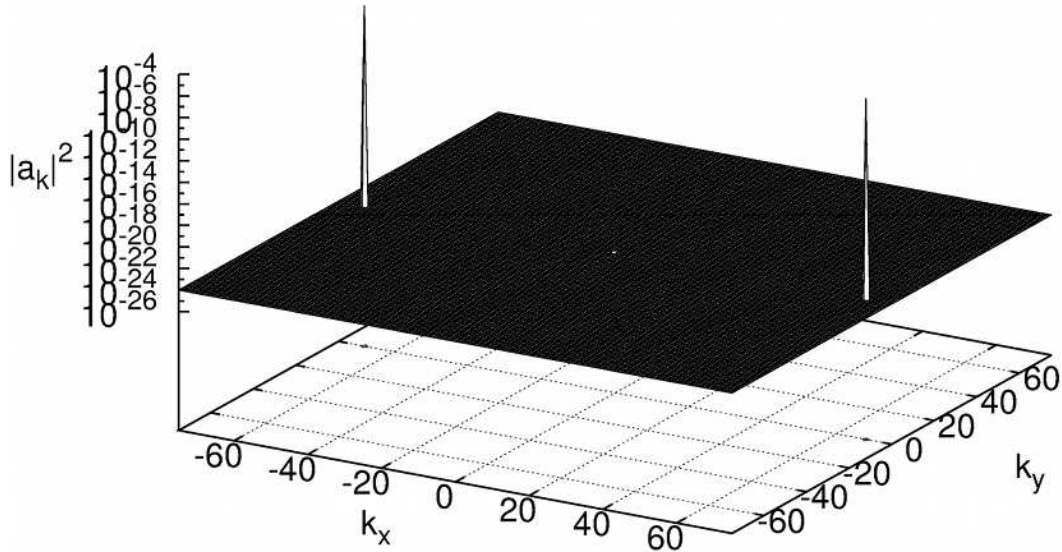

Figure 24: Instability of the standing capillary wave. Initial conditions. Time $t = 0$.

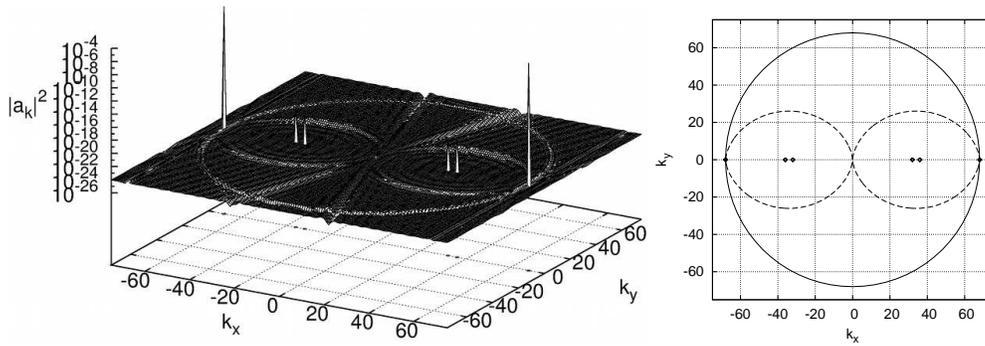

Figure 25: Instability of the standing capillary wave. Beginning of growth on the resonant curves. Some nonresonant absorption is noticeable. Time $t = 14T0$.

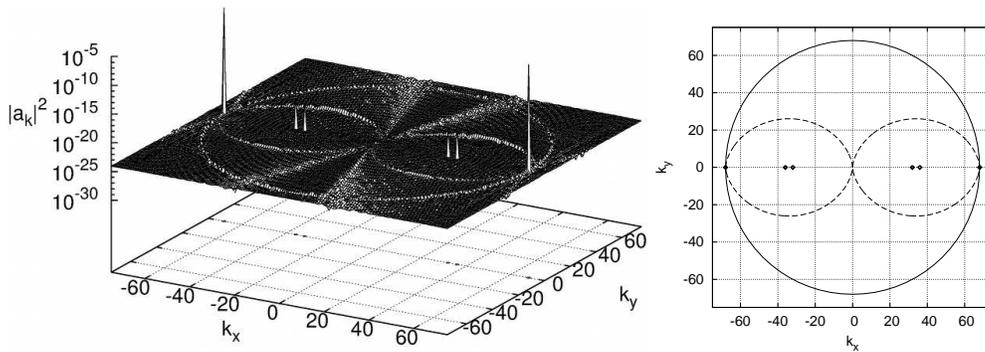

Figure 26: Instability of the standing capillary wave. Growth on resonant curves continues. Time $t = 57T0$.

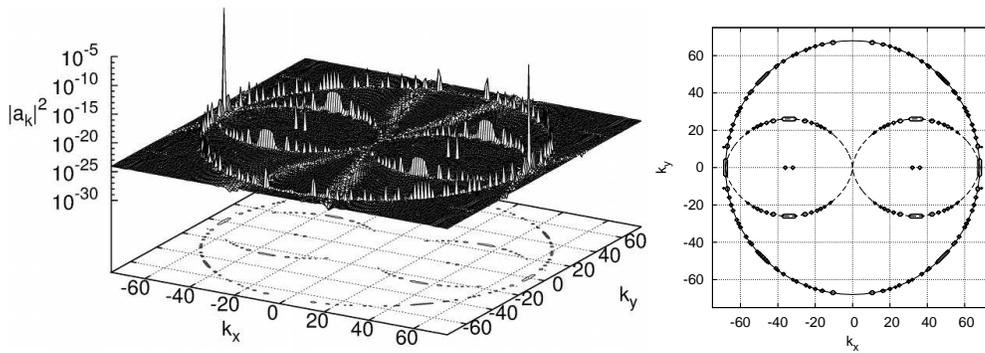

Figure 27: Instability of the standing capillary wave. Unstable harmonics are well developed. Time $t = 283T0$.

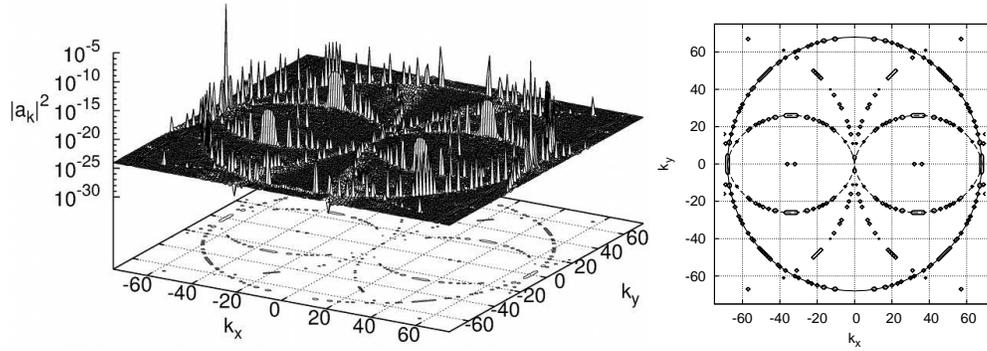

Figure 28: Instability of the standing capillary wave. One can notice formation of the forced harmonics as copies of initial circle shifted by $pm\vec{k}_0$ vectors. Time $t = 509T0$.

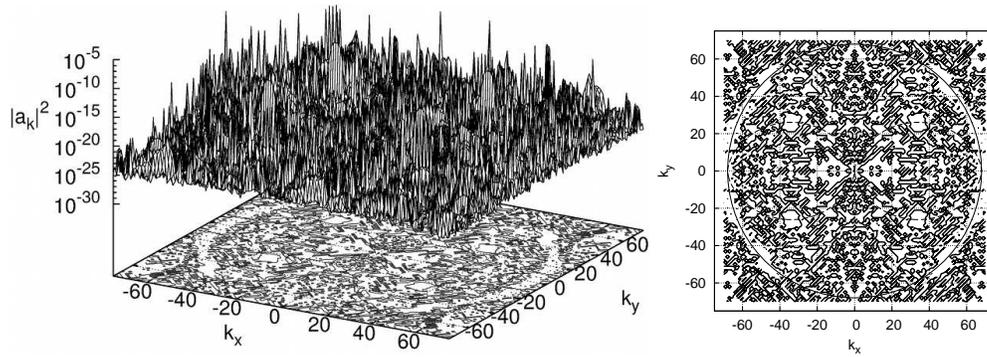

Figure 29: Instability of the standing capillary wave. Isotropization started. The plane is filled by secondary decay processes. Time $t = 1018T0$.

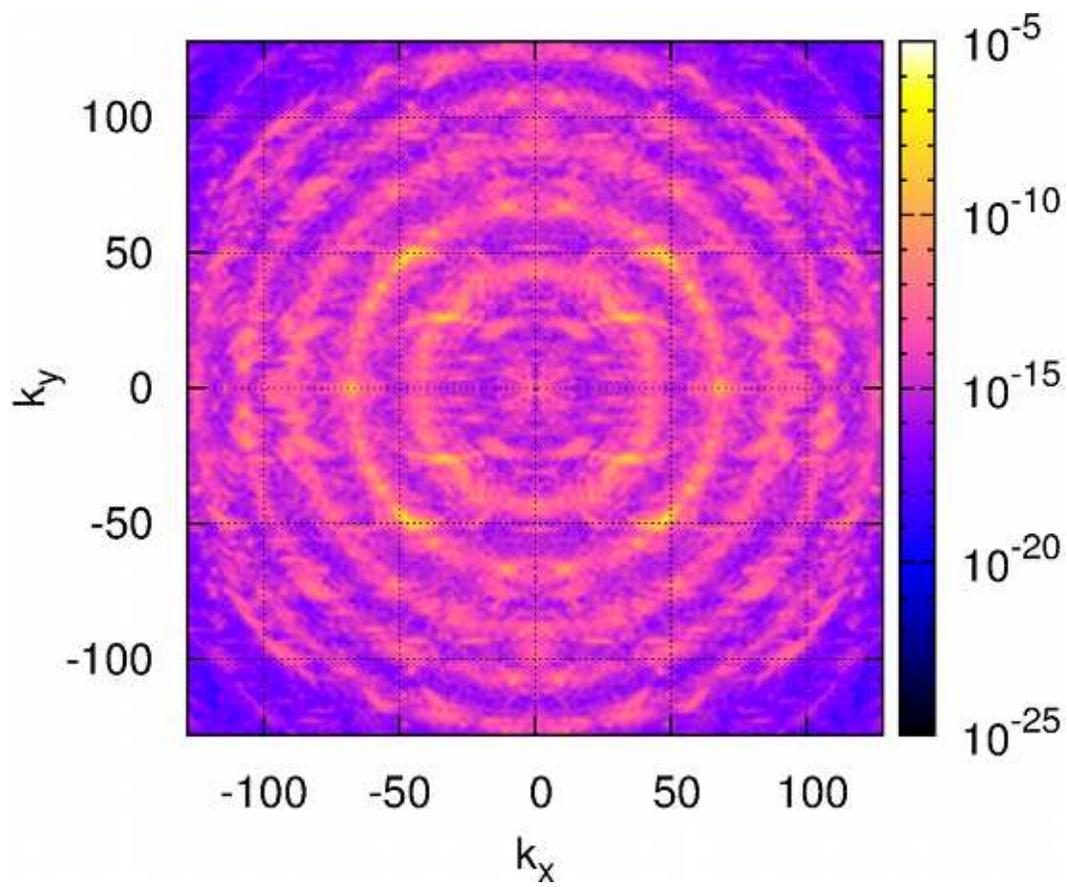

Figure 30: Instability of the standing capillary wave. Full *k*-plane. Time $t = 2587T0$.

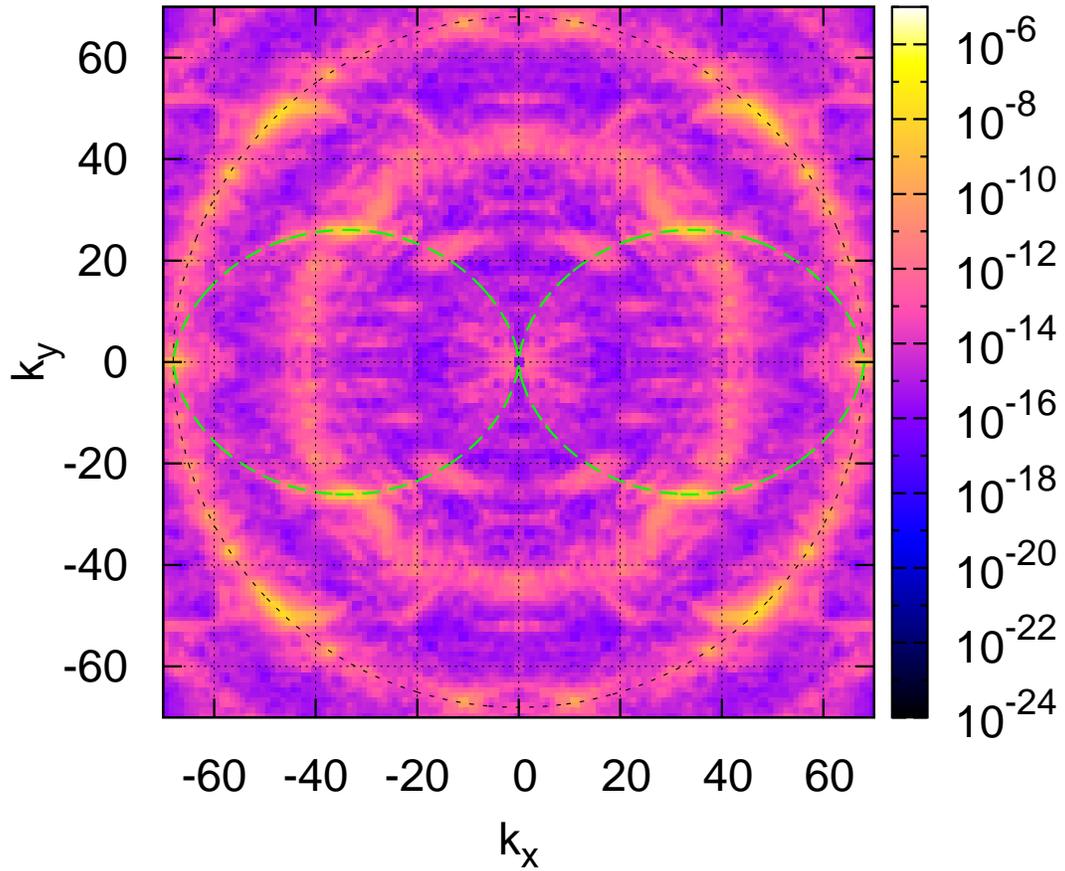

Figure 31: Instability of the standing capillary wave. Zoom of the initial instability region. The inner circle corresponds to instability of the standing waves produced in the points of the maximum of the growth rate for decay of every individual initial waves. Compare position with the most developed harmonics inside the main circle in Figure 27. Time $t = 2587T0$.

## 7.2. Standing gravity wave.

Simulation results for the standing gravity wave of steepness $\mu = 0.1$ are represented in Figures 33-37. Contour plots correspond to the level $|a_{\vec{k}}|^2 = 10^{-23}$, which is order of magnitude higher than background noise. The full picture of the $k$-plane in the final moment of simulations is represented in Figure 38. Zoom of the most interesting region of the Figure 38 (initial instability region) is represented in Figure 39. Finally we observe almost complete isotropization of the wavefield, although we started from just two waves. Weak angle dependence resembles the $\cos(\theta)$ of coupling coefficient (52). The observed process can be used for generation of isotropic wave field through initial generation of the standing wave, which in turn, through the discussed instability, will generate isotropic spectrum. This is quite nontrivial problem for direct wave generation.

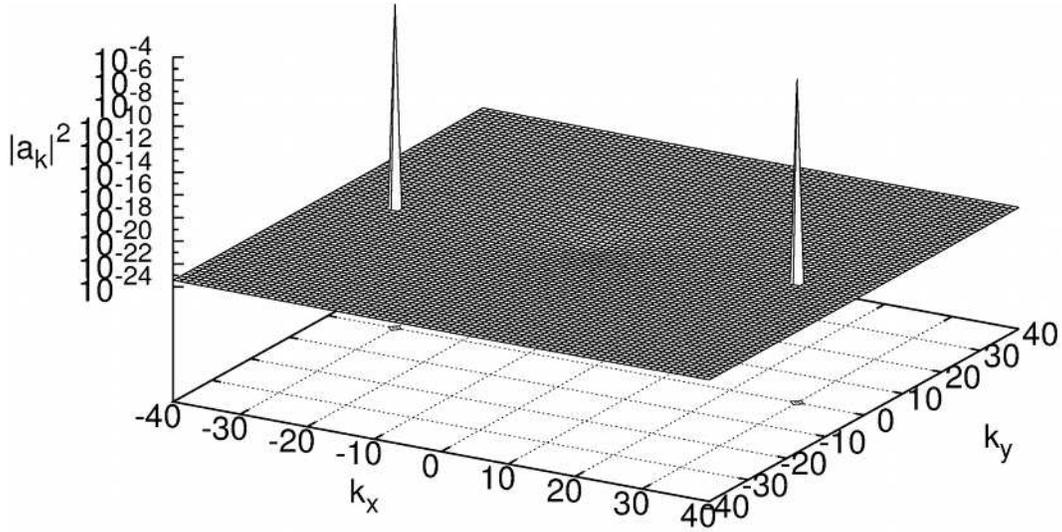

Figure 32: Instability of the standing gravity wave. Initial conditions. Time $t = 0$.

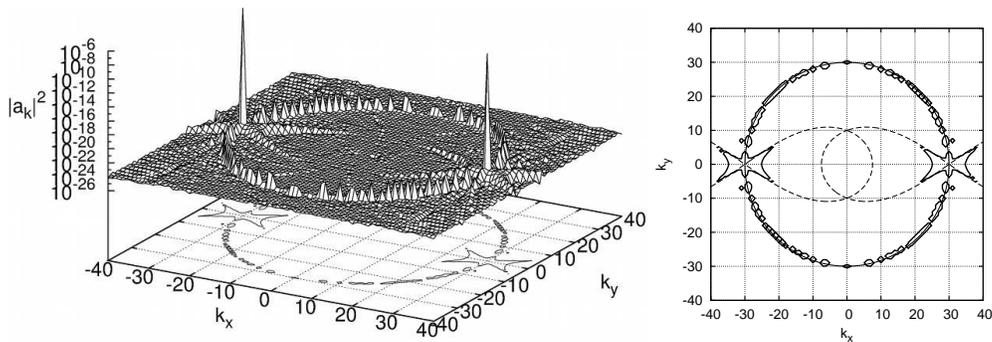

Figure 33: Instability of the standing gravity wave. Unstable harmonics on the resonant curves begin to grow. Time $t = 116T0$.

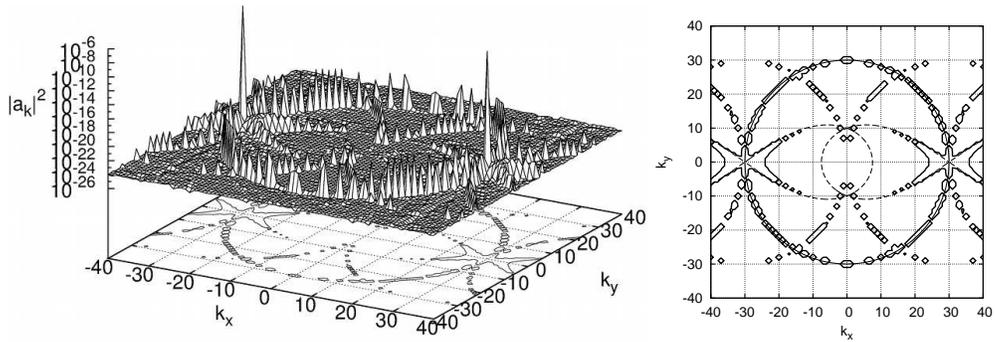

Figure 34: Instability of the standing gravity wave. Unstable harmonics are well developed. Time $t = 232T0$.

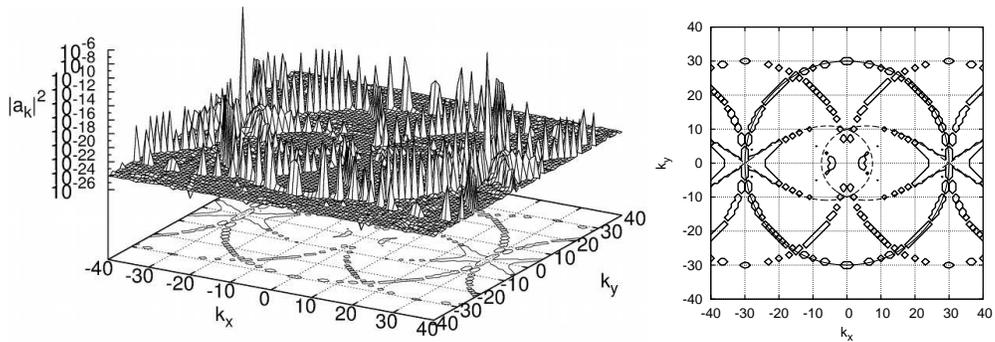

Figure 35: Instability of the standing gravity wave. Formation of forced harmonics corresponding to initial circle shifter with $\pm\vec{k}_0$ vectors. Time $t = 348T0$.

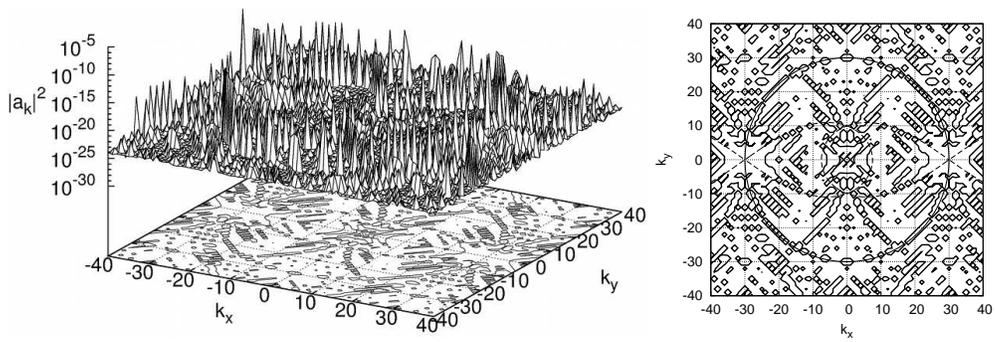

Figure 36: Instability of the standing gravity wave. Secondary processes reveal themselves. Time $t = 463T0$.

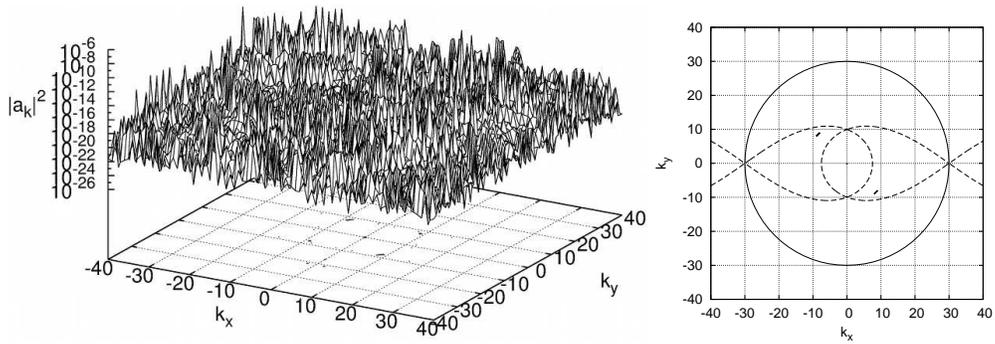

Figure 37: Instability of the standing gravity wave. Stochastization of the wavefield begins. Time $t = 580T0$.

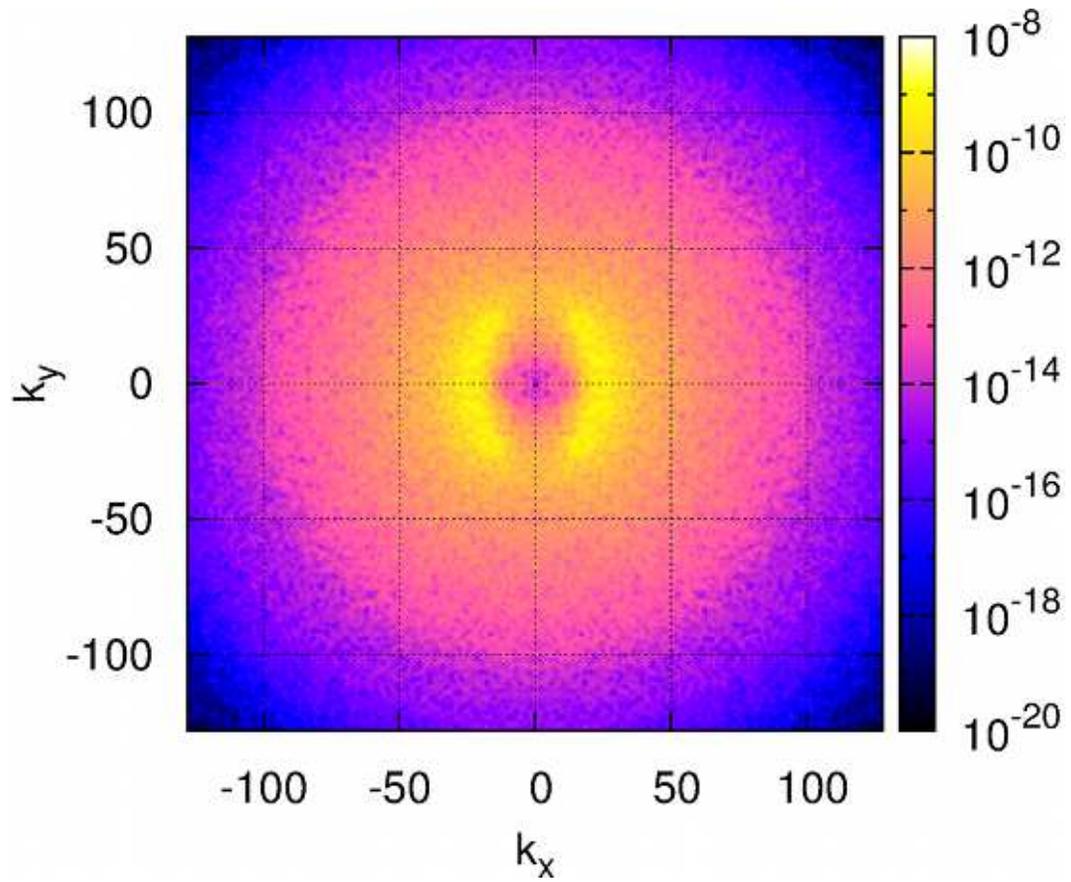

Figure 38: Instability of the standing gravity wave. Full $k$-plane. Time $t = 3068T0$.

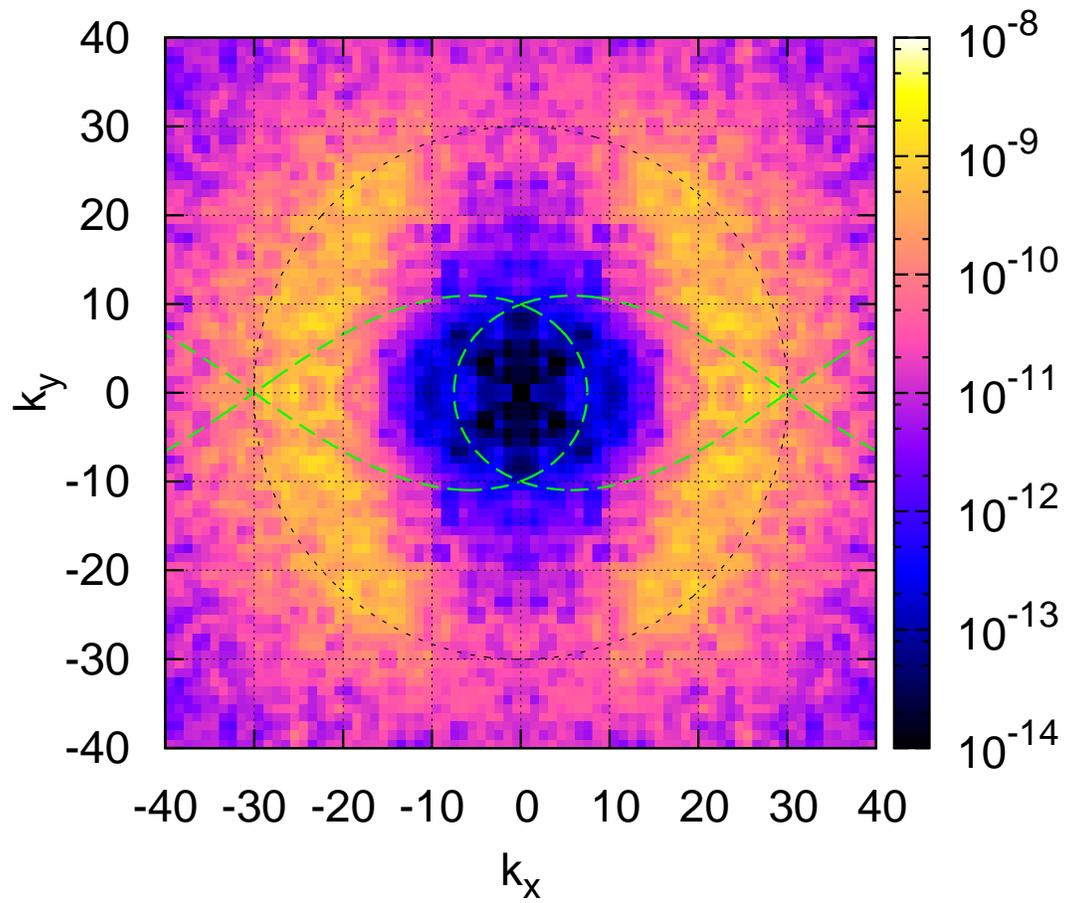

Figure 39: Instability of the standing gravity wave. Zoom of the initial instability region. Time $t = 3068T0$.

In our simulation we observed start of formation of the weakly turbulent spectrum tail (see Figure 40) and formation of Kolmogorov-Zakharov weak turbulent spectrum of direct cascade [26].

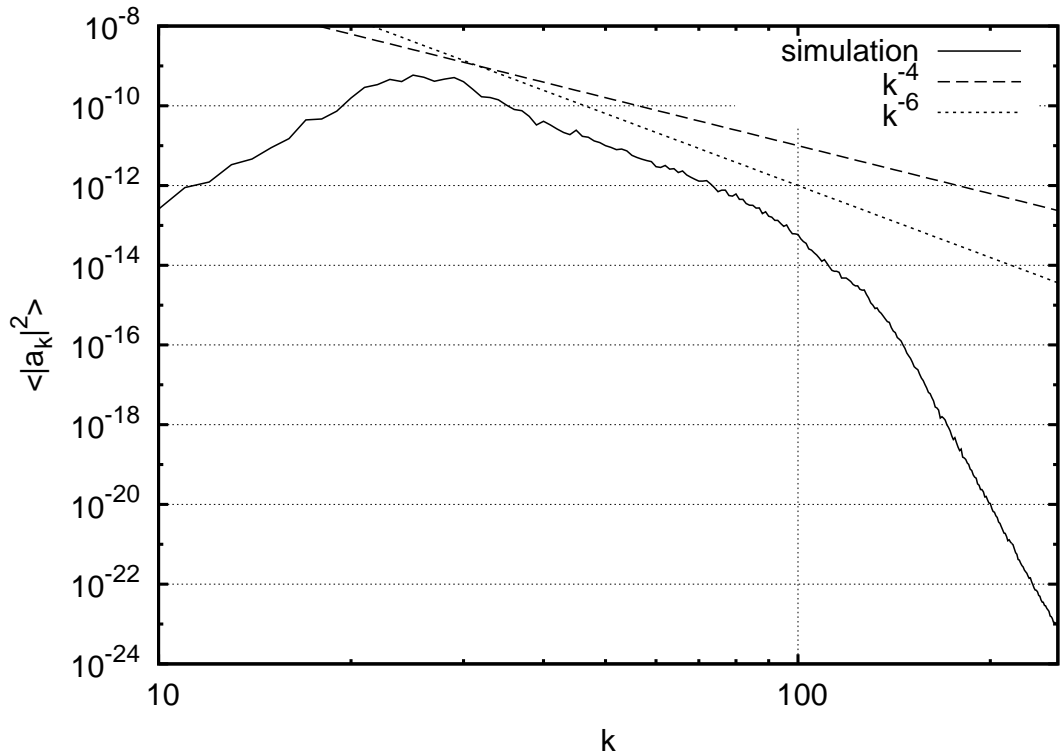

Figure 40: Instability of the standing gravity wave. Beginning of the formation of the weakly turbulent spectrum tail. Solid line: angle averaged spectrum from simulation results; dashed line: theoretically predicted KZ-spectrum [26]; dotted line: the best fit with powerlike function. Time $t = 3068 T0$.

In conclusion we have to note, that although the wave amplitudes should be high enough to make grid discreteness unimportant, at the same time they have to be low enough to satisfy weak nonlinearity conditions.

## 8. Conclusion.

We gave a complete theoretical descriptions of the three and four wave instabilities due to resonant interactions of waves. In order to simulate waves turbulence, this mechanisms have to work even on discrete grid, where exact resonance conditions are never fulfilled exactly. We demonstrated the possibility to achieve resonant interactions on discrete grid in numerical simulations. Simulation results are in good agreement with the theoretical predictions.

Also we described in details algorithm for simulation of weakly nonlinear gravity-capillary surface waves. Numerical scheme, which was used in the code, conserves Hamiltonian of the system. Features of the algorithm are used to conveniently control the adaptive time step. Described pseudo spectral method allowed us to simulate wave turbulence in numerous cases.

We discussed and simulated instability of standing wave form both capillary and gravity waves. Numerical simulations show that instability of propagating waves leads to formation of anisotropic weakly-turbulent spectra, while instability of standing waves leads to generation of almost isotropic spectra demonstrating tendency to formation of Kolmogorov-Zakharov tails. We conclude that numerical simulation of waves instability is a perfect tool for study of the wave turbulence theory. For experimental wave tanks this instability provides a very simple and robust approach which allows to produce isotropic wave field by excitation of just one standing wave.

The paper as a whole can be used as a comprehensive guide for theoretical and computational approaches to simulation of weakly nonlinear waves on the surface of the fluid.

**Acknowledgments.**

The authors gratefully wish to acknowledge the following contributions: KAO was supported by the National Science Foundation grant OCE 1131791, and during the summer visit partially supported by the grant NSh-6885.2010.2. The derivation of the condition for iterations stability and Appendix C were derived with support of the Russian Scientific Foundation grant 14-22-00259.

ZVE was partially supported by the NSF grant 1130450.

Both DAI and ZVE were supported by the Grant No. 11.G34.31.0035 of the Government of Russian Federation.

Also authors would like to thank developers of FFTW [30] and the whole GNU project [31] for developing, and supporting this useful and free software.

## Appendix A. Discrete Hamiltonian variation.

Let us derive variation of Hamiltonian ($H^{n+1} - H^n$). The $\hat{k}$ operator is self-adjoined

$$\int g\hat{k}f \, d^2r = \int f\hat{k}g \, d^2r. \tag{A.1}$$

Let us perform variation in details for quadratic part of the Hamiltonian (13) in the case of surface gravity waves.

$$H_0 = \frac{1}{2} \int (\psi \hat{k} \psi + g\eta^2) d^2r$$

$$\Delta H_0 = H_0^{n+1} - H_0^n =$$
$$= \frac{1}{2} \int (\psi^{n+1} \hat{k} \psi^{n+1} - \psi^n \hat{k} \psi^n) d^2r + \frac{g}{2} \int (\eta^{n+1^2} - \eta^{n^2}) d^2r =$$
$$= \frac{1}{2} \int (\psi^{n+1} \hat{k} \psi^{n+1} - \psi^n \hat{k} \psi^{n+1} + \psi^n \hat{k} \psi^{n+1} - \psi^n \hat{k} \psi^n) d^2r +$$
$$\frac{g}{2} \int (\eta^{n+1} - \eta^n)(\eta^{n+1} + \eta^n) d^2r = \frac{1}{2} \int \left[ (\psi^{n+1} - \psi^n) \hat{k} \psi^{n+1} + \right.$$
$$\left. + \psi^n \hat{k} (\psi^{n+1} - \psi^n) \right] d^2r + \frac{g}{2} \int (\eta^{n+1} - \eta^n)(\eta^{n+1} + \eta^n) d^2r =$$
$$= \frac{1}{2} \int (\psi^{n+1} - \psi^n) \hat{k} (\psi^{n+1} + \psi^n) d^2r + \frac{g}{2} \int (\eta^{n+1} - \eta^n)(\eta^{n+1} + \eta^n) d^2r =$$
$$= \frac{1}{2} \int \Delta \psi \hat{k} (\psi^{n+1} + \psi^n) d^2r + \frac{g}{2} \int \Delta \eta (\eta^{n+1} + \eta^n) d^2r =$$

Here and further $\Delta \psi = (\psi^{n+1} - \psi^n)$ and $\Delta \eta = (\eta^{n+1} - \eta^n)$.

Similar calculations give us all other variations.

For short let us omit integral signs in varied expressions.

Quadratic terms

$$\Delta \left( \frac{1}{2} \int \psi \hat{k} \psi d^2r \right) \longrightarrow \frac{1}{2} \Delta \psi \hat{k} \left( \psi^{n+1} + \psi^n \right); \tag{A.2}$$

$$\Delta \left( \frac{1}{2} \int \frac{\omega_k^2}{|\vec{k}|} |\eta_{\vec{k}}|^2 d\vec{k} \right) \longrightarrow \frac{1}{2} \Delta \eta_{\vec{k}} \frac{\omega_k^2}{|\vec{k}|} \left( \eta_{\vec{k}}^{n+1} + \eta_{\vec{k}}^n \right). \tag{A.3}$$

Cubic terms

$$\Delta \left( \frac{1}{2} \int \eta |\nabla \psi|^2 d^2r \right) \longrightarrow -\frac{1}{4} \Delta \psi \left( \nabla, (\eta^{n+1} + \eta^n) \nabla (\psi^{n+1} + \psi^n) \right) +$$
$$+ \frac{1}{4} \Delta \eta \left( |\nabla \psi^{n+1}|^2 + |\nabla \psi^n|^2 \right); \tag{A.4}$$

$$\Delta\left(\frac{1}{2}\int \eta\left(\hat{k}\psi\right)^2 d^2r\right) \longrightarrow -\frac{1}{4}\Delta\psi\hat{k}\left(\eta^{n+1}+\eta^n\right)\hat{k}(\psi^{n+1}+\psi^n)\right) - \qquad (A.5)$$
$$-\frac{1}{4}\Delta\eta\left((\hat{k}\psi^{n+1})^2 + (\hat{k}\psi^n)^2\right).$$

Quartic terms

$$\Delta\left(\frac{1}{2}\int \left(\eta\hat{k}\psi\right)\hat{k}\left(\eta\hat{k}\psi\right)d^2r\right) \longrightarrow \frac{1}{4}\Delta\psi\hat{k}\left[\left(\eta^{n+1}+\eta^n\right)\times\right.$$
$$\times \hat{k}\left(\eta^{n+1}\hat{k}\psi^{n+1}+\eta^n\hat{k}\psi^n\right)\right] + \qquad (A.6)$$
$$+\frac{1}{4}\Delta\eta\hat{k}\left[\left(\psi^{n+1}+\psi^n\right)\times\right.$$
$$\times \hat{k}\left(\eta^{n+1}\hat{k}\psi^{n+1}+\eta^n\hat{k}\psi^n\right)\right];$$

$$\Delta\left(\frac{1}{2}\int (\nabla^2\psi)(\hat{k}\psi)\eta^2 d^2r\right) \longrightarrow \frac{1}{8}\Delta\psi\nabla^2\left[((\eta^{n+1})^2 + (\eta^n)^2)\times\right.$$
$$\times \hat{k}(\psi^{n+1}+\psi^n)\right] +$$
$$+\frac{1}{8}\Delta\psi\hat{k}\left[((\eta^{n+1})^2 + (\eta^n)^2)\times\right. \qquad (A.7)$$
$$\times \nabla^2(\psi^{n+1}+\psi^n)\right] +$$
$$+\frac{1}{4}\Delta\eta(\eta^{n+1}+\eta^n)\times$$
$$\times(\nabla^2\psi^{n+1}\hat{k}\psi^{n+1} + \nabla^2\psi^n\hat{k}\psi^n).$$

## Appendix B. Matrix elements

We repeat formulae from [23].

$$V^{(1,2)}(\vec{k},\vec{k}_1,\vec{k}_2) = \frac{1}{4\pi\sqrt{2}}\left\{\left(\frac{A_k B_{k_1} B_{k_2}}{B_k A_{k_1} A_{k_2}}\right)^{1/4} L^{(1)}(\vec{k}_1,\vec{k}_2)\right.$$
$$\left. - \left(\frac{B_k A_{k_1} B_{k_2}}{A_k A_{k_1} A_{k_2}}\right)^{1/4} L^{(1)}(-\vec{k},\vec{k}_1) - \left(\frac{B_k B_{k_1} A_{k_2}}{A_k A_{k_1} B_{k_2}}\right)^{1/4} L^{(1)}(-\vec{k}),\vec{k}_2)\right\}, \qquad (B.1)$$

$$V^{(0,3)}(\vec{k},\vec{k}_1,\vec{k}_2) = \frac{1}{4\pi\sqrt{2}}\left\{\left(\frac{A_k B_{k_1} B_{k_2}}{B_k A_{k_1} A_{k_2}}\right)^{1/4} L^{(1)}(\vec{k}_1,\vec{k}_2)\right.$$
$$\left. + \left(\frac{B_k A_{k_1} B_{k_2}}{A_k B_{k_1} A_{k_2}}\right)^{1/4} L^{(1)}(\vec{k},\vec{k}_1) + \left(\frac{B_k B_{k_1} A_{k_2}}{A_k A_{k_1} B_{k_2}}\right)^{1/4} L^{(1)}(\vec{k}),\vec{k}_2)\right\}. \qquad (B.2)$$

$$A_k = |k|, \quad b_k = g + \sigma k^2. \qquad (B.3)$$

$$V^{(2,2)}_{\vec{k},\vec{k}_1,\vec{k}_2,\vec{k}} = 3. \tag{B.4}$$

$$a^{(0)}_{\vec{k}} = b_{\vec{k}}, \tag{B.5}$$

$$a^{(1)}_{\vec{k}} = \int \Gamma^{(1)}(\vec{k},\vec{k}_1,\vec{k}_2) b_{\vec{k}_1} b_{\vec{k}_2} \delta(\vec{k}-\vec{k}_1-\vec{k}_2)\mathrm{d}\vec{k}_1\mathrm{d}\vec{k}_2$$
$$- 2\int \Gamma^{(1)}(\vec{k}_2,\vec{k},\vec{k}_1) b^*_{\vec{k}_1} b_{\vec{k}_2} \delta(\vec{k}+\vec{k}_2-\vec{k}_2)\mathrm{d}\vec{k}_1\mathrm{d}\vec{k}_2$$
$$+ \int \Gamma^{(2)}(\vec{k},\vec{k}_1,\vec{k}_2) b^*_{\vec{k}_1} b^*_{\vec{k}_2} \delta(\vec{k}+\vec{k}_1+\vec{k}_2)\mathrm{d}\vec{k}_1\mathrm{d}\vec{k}_2, \tag{B.6}$$

$$a^{(2)}_{\vec{k}} = \int B(\vec{k},\vec{k}_1,\vec{k}_2,\vec{k}_3) b^*_{\vec{k}_1} b_{\vec{k}_2} b_{\vec{k}_3} \delta(\vec{k}-\vec{k}_1-\vec{k}_2-\vec{k}_3)\mathrm{d}\vec{k}_1\mathrm{d}\vec{k}_2\mathrm{d}\vec{k}_3 + \cdots, \tag{B.7}$$

$$\Gamma^{(1)}(\vec{k},\vec{k}_1,\vec{k}_2) = -\frac{1}{2}\frac{V^{(1,2)(\vec{k},\vec{k}_1,\vec{k}_2)}}{\omega_k - \omega_{k_1} - \omega_{k_2}}, \tag{B.8}$$

$$\Gamma^{(2)}(\vec{k},\vec{k}_1,\vec{k}_2) = -\frac{1}{2}\frac{V^{(0,3)(\vec{k},\vec{k}_1,\vec{k}_2)}}{\omega_k - \omega_{k_1} - \omega_{k_2}}, \tag{B.9}$$

$$B(\vec{k},\vec{k}_1,\vec{k}_2,\vec{k}_3) = \Gamma^{(1)}(\vec{k}_1,\vec{k}_2,\vec{k}_1-\vec{k}_2)\Gamma^{(1)}(\vec{k}_3,\vec{k},\vec{k}_3-\vec{k})$$
$$+ \Gamma^{(1)}(\vec{k}_1,\vec{k}_3,\vec{k}-\vec{k}_3)\Gamma^{(1)}(\vec{k}_2,\vec{k},\vec{k}_2-\vec{k})$$
$$- \Gamma^{(1)}(\vec{k},\vec{k}_2,\vec{k}-\vec{k}_2)\Gamma^{(1)}(\vec{k}_3,\vec{k}_1,\vec{k}_3-\vec{k}_1)$$
$$- \Gamma^{(1)}(\vec{k}_1,\vec{k}_3,\vec{k}_1-\vec{k}_3)\Gamma^{(1)}(\vec{k}_2,\vec{k}_1,\vec{k}_2-\vec{k}_1)$$
$$- \Gamma^{(1)}(\vec{k}+\vec{k}_1,\vec{k},\vec{k}_1)\Gamma^{(1)}(\vec{k}_2+\vec{k}_3,\vec{k},\vec{k}_1)$$
$$+ \Gamma^{(2)}(-\vec{k}-\vec{k}_1,\vec{k},\vec{k}_1)\Gamma^{(2)}(-\vec{k}_2-\vec{k}_3,\vec{k}_2,\vec{k}_3). \tag{B.10}$$

$$T_{1234} = \frac{1}{2}(\tilde{T}_{1234} + \tilde{T}_{2134}), \tag{B.11}$$

$$\begin{aligned}
\tilde{T}_{1234} = &-\frac{1}{16\pi^2}\frac{1}{(k_1 k_2 k_3 k_4)^{1/4}} \\
&\times \Big\{-12 k_1 k_2 k_3 k_4 - 2(\omega_1 + \omega_2)^2 [\omega_3 \omega_4 ((\vec{k}_1 \cdot \vec{k}_2) - k_1 k_2) \\
&+ \omega_1 \omega_2 ((\vec{k}_3 \cdot \vec{k}_4) - k_3 k_4)]\frac{1}{g^2} \\
&- 2(\omega_1 - \omega_3)^2 [\omega_2 \omega_4 ((\vec{k}_1 \cdot \vec{k}_3) + k_1 k_3) + \omega_1 \omega_3 ((\vec{k}_2 \cdot \vec{k}_4) + k_2 k_4)]\frac{1}{g^2} \\
&- 2(\omega_1 - \omega_4)^2 [\omega_2 \omega_3 ((\vec{k}_1 \cdot \vec{k}_4) + k_1 k_4) + \omega_1 \omega_4 ((\vec{k}_2 \cdot \vec{k}_3) + k_2 k_3)]\frac{1}{g^2} \\
&+ [(\vec{k}_1 \cdot \vec{k}_2) + k_1 k_2][(\vec{k}_3 \cdot \vec{k}_4) + k_3 k_4] \\
&+ [-(\vec{k}_1 \cdot \vec{k}_3) + k_1 k_3][-(\vec{k}_2 \cdot \vec{k}_4) + k_2 k_4] \\
&+ [-(\vec{k}_1 \cdot \vec{k}_4) + k_1 k_4][-(\vec{k}_2 \cdot \vec{k}_3) + k_2 k_3] \\
&+ 4(\omega_1 + \omega_2)^2 \frac{[(\vec{k}_1 \cdot \vec{k}_2) - k_1 k_2][-(\vec{k}_3 \cdot \vec{k}_4) - k_3 k_4]}{\omega_{1+2} - (\omega_1 + \omega_2)^2} \\
&+ 4(\omega_1 - \omega_3)^2 \frac{[(\vec{k}_1 \cdot \vec{k}_3) + k_1 k_3][(\vec{k}_2 \cdot \vec{k}_4) + k_2 k_4]}{\omega_{1-3} - (\omega_1 - \omega_3)^2} \\
&+ 4(\omega_1 - \omega_4)^2 \frac{[(\vec{k}_1 \cdot \vec{k}_4) + k_1 k_4][(\vec{k}_2 \cdot \vec{k}_3) + k_2 k_3]}{\omega_{1-4} - (\omega_1 - \omega_4)^2}\Big\}.
\end{aligned} \tag{B.12}$$

### Appendix C. On the stability of the weakly nonlinear free surface hydrodynamic model

In details this question was considered in [32] but here we shall follow original consideration which was done by A. I. Dyachenko in 1995 (result was mentioned in [33]) with some changes for 3D hydrodynamics.

Let us consider large scale solution $(\eta_0, \psi_0)$ of (19) with small scale perturbations $(\delta\eta, \delta\psi)$

$$\eta = \eta_0 + \delta\eta, \quad \psi = \psi_0 + \delta\psi. \tag{C.1}$$

After substitution of (C.1) into (19), linearization with respect to small amplitude perturbations, taking into account difference of scales, resulting in "frozen coefficients" (terms with $\eta_0$ and $\psi_0$ can be factored out from the expressions with operator $\hat{k}$ and derivatives) in the case when we keep only first nonlinear terms in (19), which resulted from cubic or three waves terms in Hamiltonian (17), yields the following dispersion relation for perturbations:

$$(\omega_k - \vec{v}_0\vec{k})^2 = \omega_{0k}^2 - (\hat{k}\psi_0)^2 k^2, \tag{C.2}$$

here we introduced $\vec{v}_0 = \vec{\nabla}\psi_0$ which is similar to velocity on the surface of the fluid and $\omega_{0k}^2 = \sigma k^3 + gk$ which is squared linear dispersion relation for gravity-capillary waves on the surface of fluid of infinite depth. Correction $\vec{v}_0\vec{k}$ in the left hand side is nothing else but Doppler shift. As one can see, for small enough scales (large enough $k$-s) the second term in the right hand side of (C.2) will prevail resulting in nonzero imaginary part for $\omega_k$. This means that we have small scale (or high frequency) instability.

If we consider the cubic terms in (19), which resulted from quartic or four waves terms in Hamiltonian (17), equation for perturbations yields (keeping only the first nonlinear terms):

$$(\omega_k - \vec{v}_0\vec{k})^2 = \omega_{0k}^2 + (\hat{k}\psi_0)(\vec{\nabla}^2\psi_0)k. \tag{C.3}$$

As one can see, the dispersion relation (C.3) has no instability at least in the first order of nonlinearity. It means that even in the case of capillary waves, where three waves interactions are the major process, presence of terms corresponding to four waves processes stabilizes the equation. If we consider the full expression for dispersion relation resulting from (19) with cubic terms:

$$(\omega_k - \vec{v}_0\vec{k})^2 = \omega_{0k}^2 + (\hat{k}\psi_0)(\vec{\nabla}^2\psi_0)k + 2(\hat{k}\psi_0)(\hat{k}[\eta_0\hat{k}\psi_0] + \eta_0\vec{\nabla}^2\psi_0)k^2 \tag{C.4}$$
$$- \left\{(\hat{k}[\eta_0\hat{k}\psi_0])^2 + 2\hat{k}[\eta_0\hat{k}\psi_0](\eta_0\vec{\nabla}^2\psi_0) + (\eta_0\vec{\nabla}^2\psi_0)^2\right\}k^2,$$

one can see, that instability reappears only with fourth order nonlinearity terms, which means that it can influence only computations with relatively high steepness or with very high values of $k$. This instability can be eliminated by proper canonical change of variables, as it was shown in [32].